\documentclass[prl,twocolumn,aps,letterpaper, preprintnumbers,superscriptaddress]{revtex4-2}

\usepackage{graphicx}
\usepackage{dcolumn}
\usepackage{bm}

\usepackage{amsthm,amsmath,amssymb}
\usepackage{mathrsfs}
\usepackage{appendix}
\usepackage{amstext} 
\usepackage{url}
\usepackage{float} 
\usepackage[usenames,dvipsnames]{color}
\usepackage[colorlinks=true,citecolor=blue,urlcolor=black]{hyperref}
\usepackage{dcolumn}
\usepackage{booktabs}
\usepackage{graphicx}
\usepackage[linesnumbered,algoruled,boxed,lined]{algorithm2e} 
\usepackage{siunitx}
\usepackage{makecell}

\newcommand{\be}{\begin{equation}}
\newcommand{\ee}{\end{equation}}

\newcommand{\sket}[1]{{\ensuremath{\lvert#1\rangle}}}
\newcommand{\lket}[1]{{\ensuremath{\left\lvert#1\right\rangle}}}
\newcommand{\ket}[1]{\if@display\lket{#1}\else\sket{#1}\fi}

\usepackage{stackengine}
\newcommand{\sbra}[1]{{\ensuremath{\langle#1\rvert}}}
\newcommand{\lbra}[1]{{\ensuremath{\left\langle#1\right\rvert}}}
\newcommand{\bra}[1]{\if@display\lbra{#1}\else\sbra{#1}\fi}

\newcommand{\sketbra}[2]{{\ensuremath{\lvert #1\rangle\!\langle #2\rvert}d}}
\newcommand{\lketbra}[2]{{\ensuremath{\left\lvert #1\right\rangle\!\!\left\langle #2\right\rvert}}}
\newcommand{\ketbra}[2]{\if@display\lketbra{#1}{#2}\else\sketbra{#1}{#2}\fi}

\newcommand{\ymx}{\textcolor{black}}

\theoremstyle{plain}

\theoremstyle{definition}

\newcommand{\lzv}{\textcolor{black}}
\newcommand{\lz}{\textcolor{black}}

\newcommand{\zy}{\textcolor{black}}
\newcommand{\hly}{\textcolor{black}}

\begin{document}
\title{Experimental Quantum Communication Overcomes the Rate-loss Limit without Global Phase Tracking}
\author{Lai Zhou}
\thanks{These authors contributed equally.}
\affiliation{Beijing Academy of Quantum Information Sciences, Beijing 100193, China} 
\author {Jinping Lin}
\thanks{These authors contributed equally.}
\affiliation{Beijing Academy of Quantum Information Sciences, Beijing 100193, China} 
\author{Yuan-Mei Xie}
\thanks{These authors contributed equally.}
\affiliation{National Laboratory of Solid State Microstructures and School of Physics, Collaborative Innovation Center of Advanced Microstructures, Nanjing University, Nanjing 210093, China}
\author{Yu-Shuo Lu}
\thanks{These authors contributed equally.}
\affiliation{National Laboratory of Solid State Microstructures and School of Physics, Collaborative Innovation Center of Advanced Microstructures, Nanjing University, Nanjing 210093, China}
\author{Yumang Jing}
\affiliation{Beijing Academy of Quantum Information Sciences, Beijing 100193, China} 
\author{Hua-Lei Yin}\email{hlyin@nju.edu.cn}
\affiliation{National Laboratory of Solid State Microstructures and School of Physics, Collaborative Innovation Center of Advanced Microstructures, Nanjing University, Nanjing 210093, China}
\affiliation{Beijing Academy of Quantum Information Sciences, Beijing 100193, China} 
\author{Zhiliang Yuan}\email{yuanzl@baqis.ac.cn}
\affiliation{Beijing Academy of Quantum Information Sciences, Beijing 100193, China} 
\date{\today}

\begin{abstract}
Secure key rate (SKR) of point-point quantum key distribution (QKD) is fundamentally bounded by the rate-loss limit. Recent breakthrough of twin-field (TF) QKD can overcome this limit and enables long distance quantum communication, 
but its implementation necessitates complex global phase tracking and requires strong phase references which not only add to noise but also reduce the duty cycle for quantum transmission.  Here, we resolve these shortcomings, and importantly achieve even higher SKRs than TF-QKD, via implementing an innovative but simpler measurement-device-independent QKD which realizes repeater-like communication through asynchronous coincidence pairing. Over 413 and 508~km optical fibers, we achieve finite-size SKRs of 590.61 and 42.64~bit/s, which are respectively 1.80 and 4.08 times of their corresponding absolute rate limits. Significantly, the SKR at 306~km exceeds 5~kbit/s and meets the bitrate requirement for live one-time-pad encryption of voice communication. Our work will bring forward economical and efficient intercity quantum-secure networks.
\end{abstract} 
 
%

\maketitle
 
\textit{Introduction.}---
Quantum key distribution (QKD)~\cite{bennett2014quantum, ekert1991quantum} has been theoretically proven secure~\cite{scarani2009security,xu2020secure
} to allow remote parties to share secret keys 
by the laws of physics. Its prospect for real-world use has motivated rapid experimental development over past forty years in terms of secure key rates (SKRs)~\cite{
yuan201810}, transmission distance ~\cite{
frohlich2017long,boaron2018secure}
and network deployment~\cite{
Peev_2009,sasaki2011field,
dynes2019cambridge,chen2021integrated}. However, realistic devices may have imperfections that could be exploited by an eavesdropper (Eve)~\cite{xu2020secure}, and among which detectors are conceivably the most vulnerable~\cite{lydersen2010hacking}. 
Fortunately, concerns on detectors have led to proposals~\cite{Braunstein2012side,lo2012measurement} of using an intermediate measurement node to close all measurement device related security loopholes. 
Additionally, the measurement node can naturally be shared by many users to form a star-type network~\cite{tang16_MDI_Network}, thus reducing the resource requirement for expensive detectors. 

\begin{figure}[t]
\centering
\includegraphics[width=8.6cm]{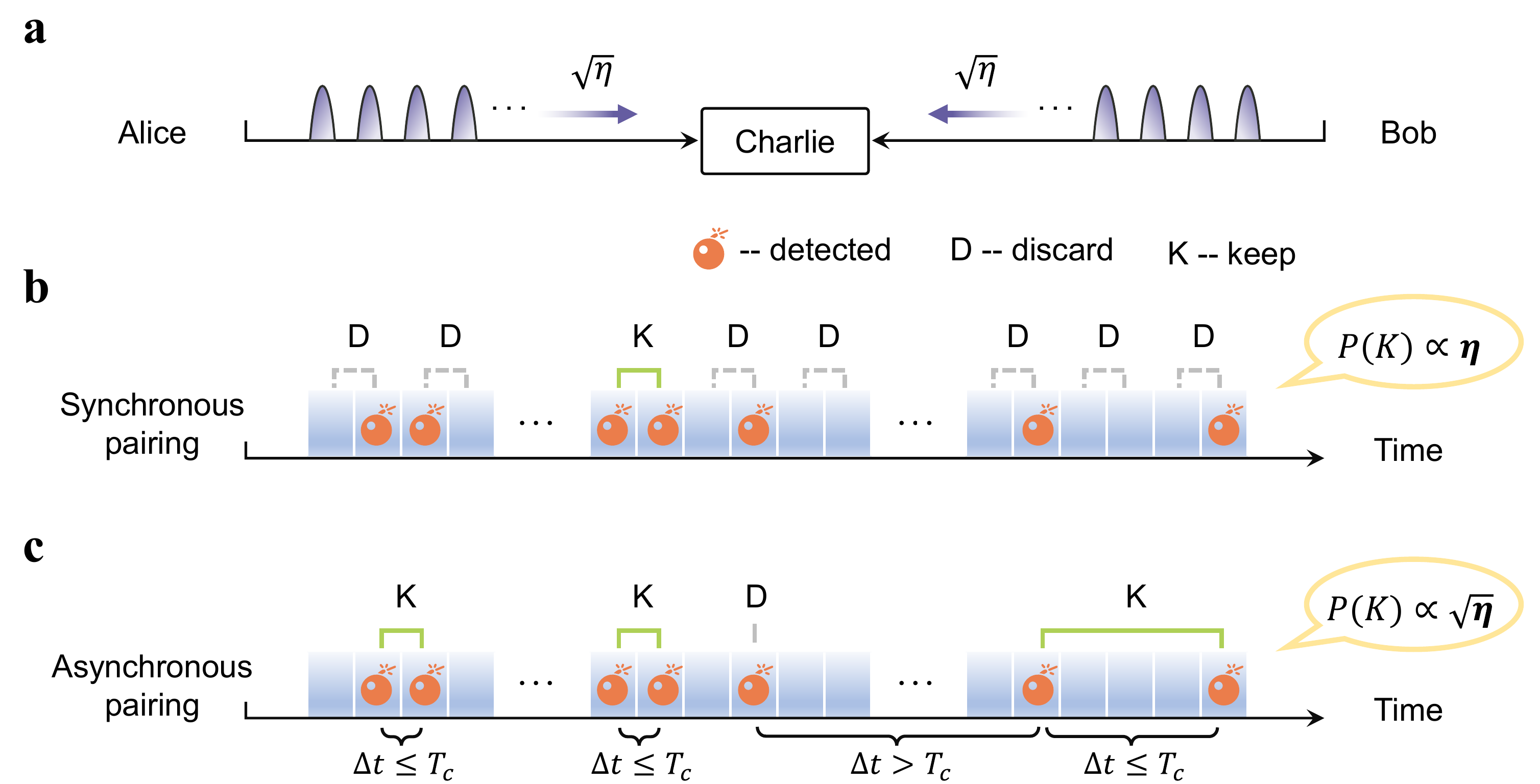}
\caption{\textbf{Schematics for MDI-QKD protocols.} \textbf{a}, Generic MDI-QKD; Alice and Bob each sends a train of encoded weak coherent pulses to the intermediate node, Charlie. 
The transmittance of the entire quantum channel is denoted as $\eta$, so each user's segment has $\sqrt{\eta}$ transmittance.
\textbf{b}, Synchronous coincidence pairing;  In original time-bin MDI-QKD, Alice and Bob apply pair-wise global phase randomization. A valid coincidence occurs when both time bins registered a photon click.  All single clicks are discarded. 
Its coincidence probability is proportional to $\eta$, \textit{i.e.}, $P(K) \propto \eta$. 
\textbf{c},  Asynchronous coincidence pairing.  In asynchronous MDI-QKD,  Alice and Bob apply independent phase slice randomization individually for each pulse. This allows innovative, post-measurement pairing of photon clicks with temporal separation within $T_c$.   
We have $P(K) \propto \sqrt{\eta}$ 
\lzv{in the high count rate limit}.}\label{fig:AMDIprotocol}
\end{figure}

On top of its security and topological advantages,  measurement-device-independent (MDI) QKD~\cite{lo2012measurement} offers substantially improved signal-to-noise ratio and hence much longer communication distances as compared to conventional QKD. This is because placing the measurement node right in the middle of a communication line effectively halves the photon transmission loss, as illustrated in Fig.~\ref{fig:AMDIprotocol}\textbf{a}.  However, the loss reduction cannot immediately translate to higher SKR as original MDI-QKD has to extract its raw key bits from two-photon coincidences.  
Using time-bin MDI-QKD as an example (Fig.~\ref{fig:AMDIprotocol}\textbf{b}), strictly synchronous pairing leads to \ymx{the probability of successful coincidence $K$} to be proportional to the total channel transmittance \ymx{$\eta$},  $P(K) \propto \eta$. Consequently, the SKR of MDI-QKD remains governed by the fundamental repeater-less limit~\cite{pirandola2009,takeoka2014fundamental,pirandola17,das2021universal}. A rigorous theorem~\cite{pirandola17} expresses this limit as $R = - \log_2(1-\eta)$~\cite{pirandola17}, which is known as the absolute repeaterless key capacity or $SKC_0$ \lzv{for a point-to-point link}. 

We note that in MDI-QKD the users' lasers are independent from each other and bear no mutual phase relationship. 
Adding the ability to track the mutual phase can convert an MDI setup to a twin-field (TF) QKD implementation~\cite{Lucamarini2018},  in which single-photon events are used for distillation of quantum keys and thus its SKR becomes repeater-like and proportional to the square-root of the channel loss ($\sqrt{\eta}$). With the help of refined protocol variants~\cite{ma2018phase,wang2018twin,yin2019measurement,lin2018simple,cui2019twin,curty2019simple,maeda2019repeaterless},  TF-QKD has been repeatedly demonstrated to overcome the $SKC_{0}$ over long fibers~\cite{
fang2020implementation,chen2020sending,clivati2022,
pittaluga21,wang22,
zhou2023quantum} and a remarkable record of 833 km for fiber transmission has been achieved~\cite{wang22}. 
However, the requirement for phase tracking has brought undesirable complexities to its implementations~\cite{
fang2020implementation,chen2020sending,clivati2022,pittaluga21,wang22,
zhou2023quantum}, all requiring service fibers to synchronize the users' lasers with one exception~\cite{zhou2023quantum} that uses optical frequency combs instead.  Moreover, it must transmit strong reference signals through the quantum channel, which reduces the effective clock frequency for quantum transmission and increases the background noise.

Recently, a new variant~\cite{xie2022breaking,zeng2022mode} of MDI-QKD has been proposed to overcome $SKC_0$ using post-measurement coincidence pairing. 
As shown in Fig.~\ref{fig:AMDIprotocol}\textbf{c}, asynchronous MDI-QKD~\cite{xie2022breaking}~\ymx{(also called mode-pairing  MDI-QKD~\cite{zeng2022mode})} allows any two photon clicks to form a legitimate coincidence provided that their time separation ($\Delta t$) is shorter than a critical interval ($T_c$), within which the users' signals maintain mutually highly coherent. The relaxation in pairing rules drastically increases the coincidence probability to $P(K)\propto  \sqrt{\eta}$ 
\lzv{in the high count rate limit, when at least two clicks on average within $T_c$}. Compared to TF-QKD, asynchronous MDI-QKD offers similar repeater-like rate-loss scaling but has the advantage of not requiring global phase tracking.

In this Letter, we demonstrate the first asynchronous MDI-QKD that overcome $SKC_0$ without resorting to global phase tracking.  We implement a 3-intensity protocol enhanced by a novel click filtering to provide security against coherent attacks in the finite-size regime. 
With fiber drift as the dominant decoherence source, our MDI-QKD system allows stable asynchronous two-photon interference over large time intervals of up to 200~$\mu$s. We obtain SKRs of 590.61 and 42.64~bit/s over fiber channels of 413.73 and 508.16~km, respectively.

\textit{Protocol.}---
Our asynchronous protocol has crucial operational differences from conventional time-bin MDI-QKD. 
Each quantum pulse requires separate phase-slice randomization, which enables asynchronous coincidence pairing after photon detection
and thus the $\sqrt{\eta}$ rate scaling. Over the initial proposal~\cite{xie2022breaking}, we have further improved the pairing strategy to intensity-independence and thus gain immunity against coherent attacks, similarly to Ref.~\cite{zeng2022mode} but with an additional filtering operation for protocol efficiency. 

We define a successful photon click as the event when one and only one detector clicked in a time bin. 
Specifically, we use $(k_{a}|k_{b})$ to denote a successful click for which Alice and Bob sent their respective pulse intensities of $k_a$ and $k_b$. 
Define $[k_{a}^{\rm tot}, k_{b}^{\rm tot}]$ as an asynchronous coincidence where the combined intensity in the two time-bins Alice (Bob) sent is $k_{a}^{\rm tot}$ ($k_{b}^{\rm tot}$). 
Let $\mu_{a(b)}$, $\nu_{a(b)}$, and $o_{a(b)}$ respectively represent Alice's (Bob's) signal, decoy and vacuum intensities,  where $\mu_{a(b)}>\nu_{a(b)}>o_{a(b)}=0$. 

Assuming authenticated public message channels, execution of the asynchronous MDI-QKD protocol follows six steps summarised below. 

Step 1 (Signal preparation and detection):~For each time bin $i=1,2,...,N$, Alice randomly prepares a weak coherent pulse $\ket{e^{\textbf{i}\theta_a}\sqrt{k_a}}$ with intensity $k_a$ and probability $p_{k_{a}}$. Thereinto, random phase $\theta_a=2\pi M_{a}/M$ with $M_{a}\in \{0,1,...,M-1\}$ and random intensity $k_a$ $\in$ $\{\mu_a, \nu_a, o_a\}$.  Likewise, Bob does the same. Alice and Bob send their optical pulses to Charlie via insecure quantum channels. 
Charlie performs the 
interference measurement and records successful clicks.
For each, he broadcasts its timestamp and the corresponding detector ($D_{L}$ or $D_{R}$) that clicked.

Step 2 (Click filtering):~For each event, Alice (Bob) announces whether she (he) applied the decoy intensity $\nu_{a}$ ($\nu_{b}$) to the pulse sent.  A simple filter is applied to discard clicks $(\mu_{a}|\nu_{b})$ and $(\nu_{a}|\mu_{b})$.  All other clicks are kept. 

Step 3 (Coincidence pairing):~For all kept clicks, Alice and Bob always pair a click with its immediate next neighbour within a time interval $T_c$ to form a successful coincidence (see post-matching algorithm in the Supplemental Material). A lone click is discarded if it failed to find a partner within $T_c$. For each coincidence, Alice (Bob) computes the total intensity $k_a^{\rm tot}$ ($k_b^{\rm tot}$) she (he) used between the two time bins.

Step 4 (Sifting): For each coincidence, Alice and Bob publish their computational result, $k_a^{\rm tot}$ or $k_b^{\rm tot}$, and the phase differences they applied between the early($e$) and late ($l$) time bins, $\varphi_{a(b)}=\theta_{a(b)}^{l}-\theta_{a(b)}^{e}$. 
They discard the data if $k_a^{\rm tot} \ge \mu_a + \nu_a$ or $k_b^{\rm tot} \ge \mu_b + \nu_b$.  For $[\mu_a, \mu_b]$ coincidence,  Alice (Bob) extracts a $\boldsymbol{Z}$-basis bit 0 if she sends $\mu_{a(b)}$ in the early (late) time bin and $o_{a(b)}$ in the late (early) bin. Otherwise, an opposite bit is extracted.
For $[2\nu_a, 2\nu_b]$ coincidence, Alice and Bob calculate the relative phase difference $\varphi_{ab}=(\varphi_{a}-\varphi_{b})\mod 2 \pi$. 
If $\varphi_{ab} = 0$ or $\pi$,  Alice and Bob extract an $\boldsymbol{X}$-basis bit 0.  However, Bob will flip his bit value if $\varphi_{ab} = 0$ and both detectors clicked or $\varphi_{ab} = \pi$ and the same detector clicked twice.  The coincidence with other phase differences is discarded.
Additionally, they group their data to different sets $\mathcal{S}_{[k_{a}^{\rm tot},k_{b}^{\rm tot}]}$ and count the corresponding number $n_{[k_{a}^{\rm tot},k_{b}^{\rm tot}]}$.

Step 5 (Parameter estimation and postprocessing):
Alice and Bob use $n_{[\mu_{a},\mu_{b}]}$ random bits from $\mathcal{S}_{[\mu_{a},\mu_{b}]}$ to form the raw key $\mathcal{Z}$ and $\mathcal{Z}'$, respectively.
The parameters $s_{11}^{z}$ and $\phi_{11}^{z}$ are the number of bits and phase error rate in $\mathcal{Z}$ where both Alice and Bob sent a single-photon state. $s_{0}^{z}$ is the number of bits in $\mathcal{Z}$ where Alice sent a vacuum state. 
By applying error correction and privacy amplification with security bound $\varepsilon_{\rm{sec}}$, the secret key rate $R$ against coherent attacks in the finite-key regime can be written as 
\begin{equation}
\begin{aligned}
R=&\frac{F}{N}\left\{\underline{s}_{0}^z+\underline{s}_{11}^z\left[1-H_2\left(\overline{\phi}_{11}^z\right)\right]-\lambda_{\rm{EC}}\right.
\\ 
&-\left.\log_2\frac{2}{\varepsilon_{\rm cor}}-2\log_2\frac{2}{\varepsilon'\hat{\varepsilon}}-2\log_2\frac{1}{2\varepsilon_{\rm PA}}\right\},\label{eq_keyrate}
\end{aligned}
\end{equation}
where $F$ is the system clock frequency, $\underline{x}$ and $\overline{x}$ are the lower and upper bounds of the observed value $x$, respectively, $\lambda_{\rm{EC}}$ is the information revealed by Alice in the error correction step,
and $H_2(x)=-x\log_2x-(1-x)\log_2(1-x)$ is the binary Shannon entropy function. $\varepsilon_{\rm cor}$, $\varepsilon_{\rm PA}$, $\varepsilon'$ and $\hat{\varepsilon}$ are security coefficients regarding the secrecy and correctness. 

We remark that filtering out clicks $(\mu_{a}|\nu_{b})$ and $(\nu_{a}|\mu_{b})$ is to increase the number of  $[\mu_{a},\mu_{b}]$ coincidence, which enables a higher SKR within 600 km fiber distance. We note that only $[\mu_{a},\mu_{b}]$ coincidence is used for extracting secret key in the present filtering method since all decoy pulses are disclosed. 


\begin{figure}[t]
\centering
\includegraphics[width=8.6cm]{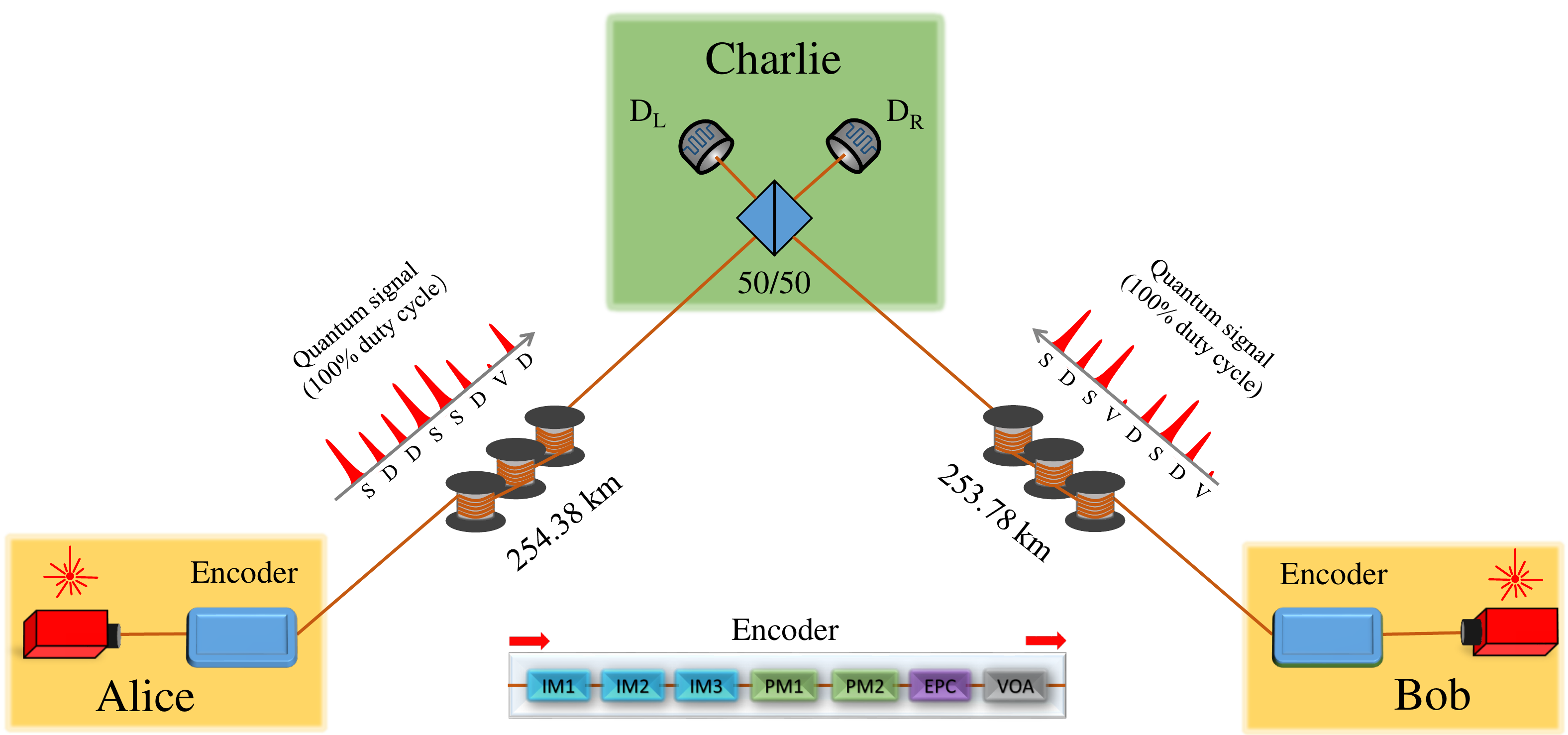}
	\caption{\lz{\textbf{Experiment setup.} Alice and Bob generate encoded weak coherent pulses with their independent ultra-stable lasers without mutual phase tracking.  The encoder box contains three intensity and two phase modulators.  
 IM: Intensity Modulator; PM: phase modulator; EPC: electrically driven polarization controller; VOA: variable optical attenuator; S: signal intensity; D: decoy intensity; V: vacuum intensity.}
	}\label{fig_experiment_setup}
\end{figure}

\textit{{Setup}.}---Our experimental setup (Fig.~\ref{fig_experiment_setup}) consists of three main modules: the senders Alice/Bob and the measurement node Charlie. Each sender contains a continuous-wave laser with a central wavelength of 1550.12~nm and featuring a short-term linewidth of 1~Hz. Their wavelengths are independently stabilized onto a transmission mode of their own high-fineness cavities using Pound-Drever-Hall (PDH) technique. An electro-optical modulator (EOM) is present in the PDH locking path so as to shift the laser frequency with respect to the cavity mode. This allows control of the lasers' mutual frequency offset by finely adjusting the radio-frequency (RF) driving frequency to the EOM.

Passing through the encoder box, the laser signal is first carved into a train of 300~ps pulses at 1~ns intervals, followed by further intensity and phase encoding according to the requirements by the asynchronous MDI-QKD protocol. 
The encoded pulses are attenuated to the single-photon level before entering their respective quantum link. The quantum channel is formed by ultra-low-loss fiber spools (G654.C ULL) with a typical loss coefficient ranging from 0.158~dB~km$^{-1}$ to 0.162~dB~km$^{-1}$. 

After pre-compensating the polarization drift, the quantum signals from Alice and Bob travel through the corresponding link segment and arrive with identical polarization at Charlie's 50/50 beam splitter for interference. The interference outcomes are detected by two superconducting nanowire single photon detectors ($D_L$ and $D_R$) having respective detection efficiencies of $78.1\%$ and $77.0\%$, dark count rates of 10.1 and 12.7~Hz \lzv{and a time jitter of about 40 ps.} Detection events are recorded by a time tagger with 300~ps gate width, and subsequently  post-processed to extract MDI-QKD protocol parameters. \textcolor{black}{Between the senders and Charlie, we use electrical signals for clock synchronization, which can be upgraded to optical synchronization as routinely used in QKD field trials~\cite{
sasaki2011field,dynes2019cambridge,chen2021integrated}.}

As compared with TF-QKD implementations~\cite{zhou2023quantum,pittaluga21,chen2022quantum,wang22}, our MDI-QKD setup is substantially simpler as it does not require optical frequency dissemination, global phase tracking, and strong phase reference signals.  With quantum transmission at 
\lzv{$100 \%$} duty cycle,  the asynchronous MDI-QKD can therefore surpass the SKR performance of TF-QKD systems~\cite{pittaluga21,zhou2023quantum} operating at the same clock frequency.

\textit{Experimental results.}---It is crucial to have high visibility interference between the users' lasers, as the visibility is the key parameter for foiling Eve's attacks that break coherence among pulses.  Theoretically,  the first-order interference between two independent lasers can reach a temporal visibility  ($V_1$) of 1 and the corresponding second-order coincidence  interference has a maximum dip visibility ($V_2$) of 0.5.  
Here, we verify our experimental setup by transmitting the pulse-carved signals over short fibers and variable optical attenuators (VOAs) and obtain the respective visibilities of $V_1 = 0.989$  and $V_2=0.484$.   
Nevertheless, the measured $V_2$ is sufficient to give an $\boldsymbol{X}$-basis QBER ($E_{x}$) of 0.26, which has a theoretical minimum of 0.25.  

\begin{figure}[t]
\centering
\includegraphics[width=8.6cm]{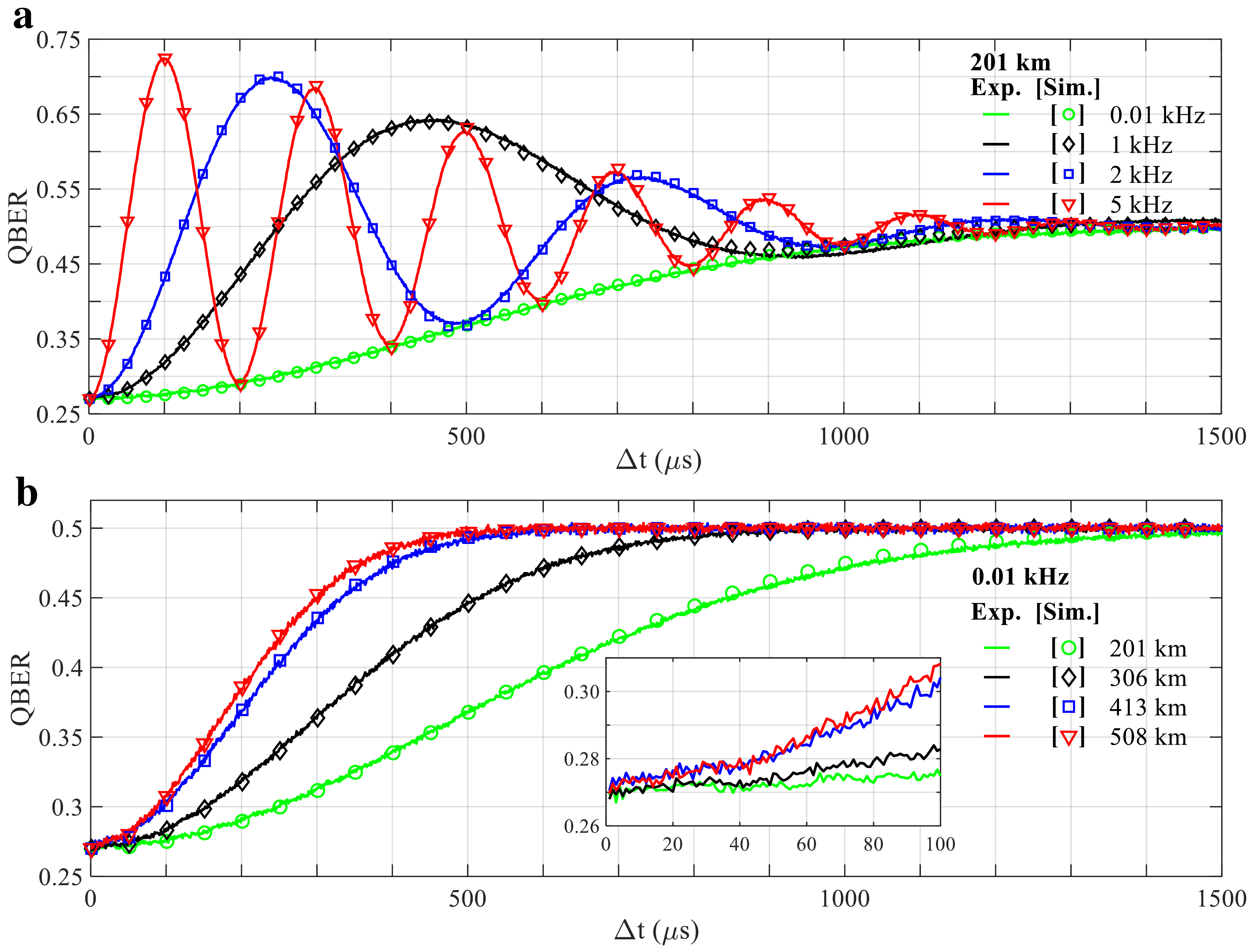}
	\caption{\lz{\textbf{Characterization of the asynchronous two-photon interference.} \textbf{a}, Evolution of the $\boldsymbol{X}$-basis QBER for different laser frequency offsets for a fixed fiber distance of 201.86~km.  \textbf{b}, $\boldsymbol{X}$-basis QBER as a function of time interval for different fiber distances, ranging from $201.86$ to $508.16$ km.} Inset: magnified view for the region between 0 and 100~$\mu$s.}\label{HOM}
\end{figure}

We then verify the effect by the lasers' mutual frequency offset ($\Delta \lzv{f}$) and the fiber length fluctuation on the $\boldsymbol{X}$-basis QBER. We run several asynchronous MDI-QKD experiments over the quantum channel of 201.86~km fibers while setting different offsets of $<0.01$, $1$, $2$ and $5$~kHz.  Here,  signal modulation is performed exactly as the protocol prescribes, including the 16-slices phase randomization.  
We collect 5~s data for each $\Delta \lzv{f}$.  For convenience, we extract the $\boldsymbol{X}$-basis QBER from $[2\mu_a, 2\mu_b]$  coincidences, \textit{i.e.}, among click events when both Alice and Bob transmitted a signal state ($\mu_a$ or $\mu_b$).  A correct coincidence corresponds to either Alice and Bob used identical phase difference 
(\hly{$\varphi_{ab}=0$}) for the two time-bins and the same detector clicked twice or 
$\hly{\varphi_{ab}}=\pi$ and each detector clicked once.
The experimental result is plotted in Fig.~\ref{HOM}\textbf{a}. 
With a negligible offset of 0.01~kHz, the fiber fluctuation dominates the dephasing between two coincident time-bins, leading to a monotonous increase of $\boldsymbol{X}$-QBER to 0.5 when the time separation reaches \zy{1.5}~$\mu$s. In the presence of a larger frequency offset, $\boldsymbol{X}$-QBER exhibits oscillations at the corresponding offset frequency with a damping amplitude. The minimas are bounded by the green curve ($\Delta\lzv{f}$ $< $10~Hz).
The oscillation is due to the mutual phase evolution of the two lasers.   

The effect by fiber fluctuation and frequency offset on the asynchronous two-photon interference can be theoretically derived as (see the Supplemental Material)
\begin{equation}
    E_x = \frac{1-V_2}{2} + \frac{V_2}{2}\left[1-e^{-\sigma^2 \Delta t^2 / 2} {\rm cos}( 2 \pi \Delta \lzv{f} \Delta t)\right],
\label{eq:sim_theory}
\end{equation}
\noindent where 
$\sigma$ is the standard deviation of the fiber drift rate. 
This equation can near perfectly reproduce the experimental results, as shown in Fig.~\ref{HOM}\textbf{a}.  Here, we set $V_2 = 0.46$ taking into account further deterioration by phase randomization error and use an empirical value of 2100~rad s$^{-1}$ for fiber phase drift~\cite{Lucamarini2018}.  No other fitting parameters are used.

Hundreds of kilometers fiber can contribute a phase drift rate of several kHz, and this will limit the longest interval within which two clicks can be paired with an acceptable error ratio.  At $\Delta\lzv{f}$  $< 10$~Hz, it is possible to achieve $\boldsymbol{X}$-QBER of less than 0.30 over a coincidence interval of 200~$\mu$s, as shown in Fig.~\ref{HOM}\textbf{a}.  To evaluate for longer distances,  we keep the $\Delta\lzv{f}$ below 10 Hz and then measure $\boldsymbol{X}$-basis QBER over different fiber lengths. As shown in Fig.~\ref{HOM}\textbf{b}, the $\boldsymbol{X}$-QBER deteriorates faster for longer fibers. 
At the maximum length of 508~km, the $\boldsymbol{X}$-QBER reaches 0.30 at the time interval of 85~$\mu$s, see Fig.~\ref{HOM}\textbf{b}, inset. 
However, the actual average interval is much smaller and we can therefore expect lower $\boldsymbol{X}$-QBER because our scheme pairs just adjacent photon clicks.
As demonstrated later, we are able to use a large $T_c$ of 200~$\mu$s for 508~km while still obtaining an acceptable $\boldsymbol{X}$-QBER of 0.293.
We perform the theoretical simulation using Eq.~\ref{eq:sim_theory} and find excellent agreement with experiments, see Fig.~\ref{HOM}\textbf{b}.  In the simulation, we use empirical drift rates of $2100$, $3400$, $5300$ and $5900$~rad s$^{-1}$ for 201, 306, 413 to 508~km of fibers, respectively.  The used drift rates are in good agreement with previous experimental observations~\cite{Lucamarini2018,zhou2023quantum}.

Finally, we run the asynchronous MDI-QKD protocol over for four fiber distances. \lzv{We globally optimize the parameters $\mu$, $\nu$, $p_\mu$, $p_\nu$ in the respective ranges [0.3, 0.5], [0, 0.1], [0.1, 0.4] and [0.1, 0.4] for a maximal secure key rate.} 
To ensure high visibility interference, we compensate for polarization and temporal drift of the photon arrivals at Charlie's 50/50 beam-splitter. 
To mitigate finite-size effects, we increase the number of sent pulses from  $4.30 \times 10^{12}$ to $7.24 \times 10^{13}$ when the fiber length increases from $201.86$ to $508.16$~km. 
The average pairing intervals are 0.43 $\mu$s and 70.89 $\mu$s when setting the $T_{c}$ as 5~$\mu$s and 200~$\mathrm{\mu}$s for 201.86 and 508.16 km fibers. We measure $\boldsymbol{X}$-basis QBER's for coincidences $[2\nu_a, 2\nu_b]$ to be $0.269$ and $0.293$, respectively. 
The detailed encoding parameters and experimental results are summarised in the Supplemental Material. 

\begin{figure}[t]
\centering
\includegraphics[width=8.6cm]{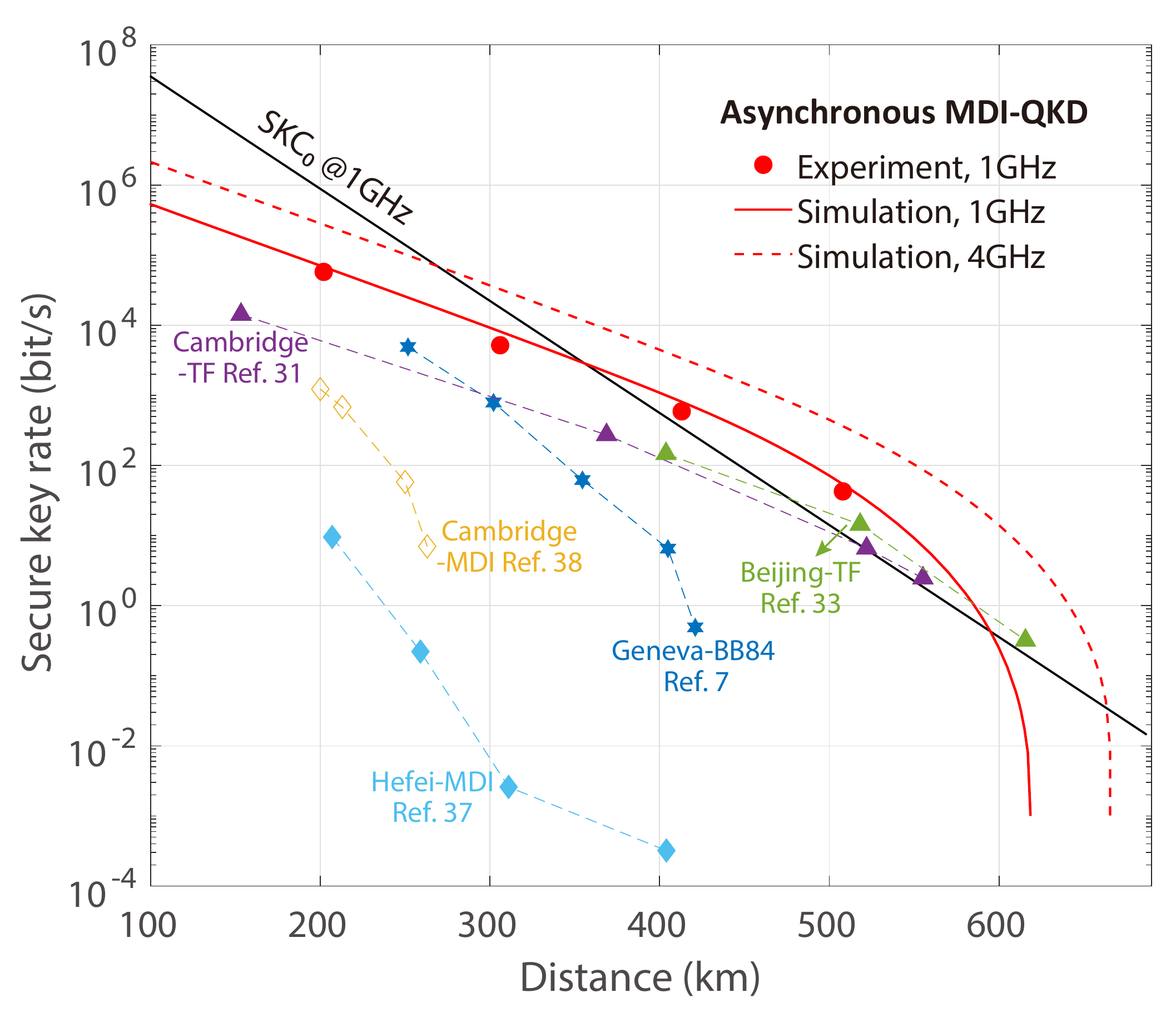}
	\caption{\textbf{Key rate simulations and results.} 
	Experimental data (solid red circles) and its simulation  (solid red line) are plotted together with the absolute repeaterless bound, $\rm{SKC}_0$. The red curve is the simulation result with the experimental
    parameters and a total of $7.24\times 10^{13}$ transmitted quantum pulses.  Simulation for 4~GHz clock rate (red dashed line) is also shown. In the simulation, we set empirical drift rates of 5900 rad s$^{-1}$.
	Results of state-of-the-art QKD experiments~\cite{yin2016measurement,boaron2018secure,woodward2021gigahertz,pittaluga21,zhou2023quantum} are included for comparison. }\label{fig:SKR}
\end{figure}

Figure~\ref{fig:SKR} presents our experimental results (red circle) in terms of SKR versus fiber distance, together with the theoretical simulation (solid red line). We include also the absolute $SKC_0$ to prove the repeater-like behavior for our system. Taking into account of finite-size effects, we obtain SKRs of 57.63k, 5.18k, 590.61 and 42.64 bit/s for $201.86$, $306.31$, $413.73$ and $508.16$ km, respectively. Remarkably, the SKRs at $413.73$ and $508.16$~km beat their respective linear bounds with considerable margins, being $1.80$ and $4.08$ times higher than $SKC_0$. This is the first time for a QKD system to beat $SKC_0$ without resorting to complex global phase tracking.

To further appreciate the progress made by our system and protocol, we compare our results with the state-of-the-art QKD systems.  
Our asynchronous system has absolute advantage over existing MDI-QKD systems~\cite{yin2016measurement,woodward2021gigahertz} 
implementing synchronous coincidence pairing. Its repeater-like rate scaling allows it to beat the QKD system~\cite{boaron2018secure} operating at a higher clock rate of 2.5~GHz.  
Strikingly, our system achieves higher performance even than TF-QKD systems~\cite{pittaluga21,zhou2023quantum} operating at the identical clock of 1~GHz over the distances between 200 and 500~km, despite that our system is substantially simpler and does not need global phase reconciliation.   
Over longer distances, our system performance is restricted by the loss of coincidence pairing efficiency due to both shortened $T_c$ and less frequent photon clicks.  This problem can be relieved partially by increasing the clock rate as simulated (dashed line, Fig.~\ref{fig:SKR}).   Alternatively, we may use side-band stabilization technique~\cite{zhou2023quantum} that can reduce the fiber drift rate by a factor of 1000 and thus substantially enlarges $T_c$ to the order of 100~ms.

\textit{Discussion.}---We have realized the first MDI-QKD experiment that breaks the $SKC_0$ bound, extending the maximal distance from 404 km to 508 km and improving the SKR over 400 km by more than 6 orders of magnitude.
This success is attributed to the asynchronous pairing method and the optimisation strategy through click filtering.
The removal of phase tracking ensures an economical and efficient inter-city quantum-secure network. In the future, we expect to improve the clock rate and also use less-demanding lasers for additional practicality enhancement. Increasing the system clock rate can proportionally shorten the pairing intervals, thereby further reducing the error rate and improving noise tolerance. 
We believe that the asynchronous MDI-QKD experiment design can be useful for applications such as quantum repeaters, entanglement swapping and quantum internet.

\bigskip
This work was supported by the National Natural Science Foundation of China (62105034, 12274223, 62250710162), the Natural Science Foundation of Jiangsu Province (BK20211145), the Fundamental Research Funds for the Central Universities (020414380182), and the Key Research and Development Program of Nanjing Jiangbei New Aera (ZDYD20210101).

\textit{Note added.}---We note that related experimental work has been reported in Ref.~\cite{zhu2022experimental}. Both our work and Ref.~\cite{zhu2022experimental} \zy{implement} MDI-QKD with \zy{post-measurement} coincidence pairing. However, we realize the first MDI-QKD experiment that breaks the repeaterless bound and extend\zy{s} the maximal \zy{MDI-QKD} distance from 404 km to 508 km, 
while Ref.~\cite{zhu2022experimental} \zy{achieved} a maximal distance of 407~km \zy{but did not} break the repeaterless bound.


\bibliographystyle{modified-apsrev4-2_new}
\bibliography{AMDIbib.bib}
\end{document}


\title{Supplemental Material}


\author{Lai~Zhou}
\thanks{These authors contributed equally.}
\affiliation{Beijing Academy of Quantum Information Sciences, Beijing 100193, China} 
\author {Jinping~Lin}
\thanks{These authors contributed equally.}
\affiliation{Beijing Academy of Quantum Information Sciences, Beijing 100193, China} 
\author{Yuan-Mei~Xie}
\thanks{These authors contributed equally.}
\affiliation{National Laboratory of Solid State Microstructures and School of Physics, Collaborative Innovation Center of Advanced Microstructures, Nanjing University, Nanjing 210093, China}
\author{Yu-Shuo~Lu}
\thanks{These authors contributed equally.}
\affiliation{National Laboratory of Solid State Microstructures and School of Physics, Collaborative Innovation Center of Advanced Microstructures, Nanjing University, Nanjing 210093, China}
\author{Yumang~Jing}
\affiliation{Beijing Academy of Quantum Information Sciences, Beijing 100193, China} 
\author{Hua-Lei~Yin}\email{hlyin@nju.edu.cn}
\affiliation{National Laboratory of Solid State Microstructures and School of Physics, Collaborative Innovation Center of Advanced Microstructures, Nanjing University, Nanjing 210093, China}
\affiliation{Beijing Academy of Quantum Information Sciences, Beijing 100193, China} 
\author{Zhiliang~Yuan}\email{yuanzl@baqis.ac.cn}
\affiliation{Beijing Academy of Quantum Information Sciences, Beijing 100193, China} 

\maketitle

\tableofcontents

\section{Post-Matching Algorithm}
Here, we introduce the algorithm of post-matching for asynchronous measurement-device-independent quantum key distribution (MDI-QKD). In Algorithms 1 and 2 we consider the cases with and without click filtering, respectively. 
\begin{algorithm} 
\DontPrintSemicolon
\SetAlgoLined
\caption{Post-matching algorithm for the asynchronous MDI-QKD with click filtering} 
\KwIn{The time bin sequence of the click events ($D$); For each time bin $i$,
the intensity $k_a^i$, $k_b^i$ and the phase $\theta_a^{i}$, $\theta_b^{i}$; the number $N_{T_{c}}=FT_{c}$ of pulses sent within time interval $T_c$ and system clock frequency $F$.}
\KwOut{ The set  $\mathcal{S}_{[k_a^{\rm{tot}},k_b^{\rm{tot}}]}$ of paired coincidence  $[k_a^{\rm{tot}},k_b^{\rm{tot}}]$ and $n_{[k_a^{\rm{tot}},k_b^{\rm{tot}}]}$ (the total amount of data in set $\mathcal{S}_{[k_a^{\rm{tot}},k_b^{\rm{tot}}]}$).   
} 
\BlankLine
Initialize the  set  $\mathcal{S}_{[k_a^{\rm{tot}},k_b^{\rm{tot}}]}$\;
\For{$i=1:\rm{length}($D$)$~~~~~\tcp{Iterate over the detection event sequence}}
{\If{ $( k_a^i|k_b^i) = (\nu_a|\mu_b)$ \rm{or} $( k_a^i|k_b^i)= ( \mu_a|\nu_b)$~~~~~} 
{
Remove $D^i$ from $D$.~~~~~\tcp{Remove all $( \nu_a|\mu_b)$ and $(\mu_a|\nu_b)$ events from the detection sequence.}
}
}
{\For{$i = 1: \rm{length}($D$)$}
{  

\If{$D^{i+1}-D^i\le N_{T_{c}}$ \rm{and $i$-th click event have not been matched}} 
{   
\tcp{Alice and Bob obtain a pairing event}
\tcp{Calculate the combined intensity of two pairing time bins}
$k_a^{i}+k_a^{i+1} = k_a^{\rm tot}$, $k_b^{i}+k_b^{i+1} = k_b^{\rm tot}$\;
\uIf{$k_a^{\rm tot} \ge \mu_a+\nu_a$ {\rm or} $k_b^{\rm tot} \ge \mu_b+\nu_b$}
{break\;}
\uElseIf{$k_a^{\rm tot}$ =  $\mu_a$ {\rm and} $k_b^{\rm tot}$ =  $\mu_b$}{
Append $[D^{i}, D^{i+1}]$ to $\mathcal{S}_{[\mu_a,\mu_b]}$~~~~~\tcp{ assign to the coincidence $[\mu_a,\mu_b]$}

}
\uElseIf{ $k_a^{\rm tot}$ =  $2\nu_a$ {\rm and} $k_b^{\rm tot}$ =  $2\nu_b$}{
\If{ $\left(\theta_a^{i+1}-\theta_a^{i}-\theta_b^{i+1}+\theta_b^{i}\right) \mod 2\pi$ = {\rm 0} \rm{or} $\pi$
}
{Append $[D^{i}, D^{i+1}]$ to $\mathcal{S}_{[2\nu_a,2\nu_b]}$~~~~~\tcp{assign to the coincidence $[2\nu_a,2\nu_b]$}
}
}
\Else{Append $[D^{i}, D^{i+1}]$ to $\mathcal{S}_{[k_a^{\rm tot},k_b^{\rm tot}]}$~~~~~\tcp{ assign to  the coincidence $[k_a^{\rm tot},k_b^{\rm tot}]$}
}
}
}
}
\end{algorithm}

\begin{algorithm} 
\DontPrintSemicolon
\SetAlgoLined
\caption{Post-matching algorithm for the asynchronous MDI-QKD without click filtering} 
\KwIn{The time bin sequence of the click events ($D$); For each time bin $i$,
the intensity $k_a^i$, $k_b^i$ and the phase $\theta_a^{i}$, $\theta_b^{i}$; the number $N_{T_{c}}$ of pulses sent within time interval $T_c$ and system clock frequency $F$.}
\KwOut{ The set  $\mathcal{S}_{[k_a^{\rm{tot}},k_b^{\rm{tot}}]}$ of coincidence  $[k_a^{\rm{tot}},k_b^{\rm{tot}}]$ and $n_{[k_a^{\rm{tot}},k_b^{\rm{tot}}]}$ (the total amount of data in set $\mathcal{S}_{[k_a^{\rm{tot}},k_b^{\rm{tot}}]}$). 
} 
\BlankLine
Initialize the  set  $\mathcal{S}_{[k_a^{\rm{tot}},k_b^{\rm{tot}}]}$ \;

{\For{$i = 1: \rm{length}($D$)$}
{  

\If{$D^{i+1}-D^i\le N_{T_{c}}$ \rm{and $i$-th click event have not been matched}} 
{   
\tcp{Alice and Bob obtain a pairing event}
\tcp{Calculate the combined intensity of two pairing time bins}
$k_a^{i}+k_a^{i+1} = k_a^{\rm tot}$, $k_b^{i}+k_b^{i+1} = k_b^{\rm tot}$\;
\uIf{$k_a^{\rm tot} \ge \mu_a+\nu_a$ {\rm or} $k_b^{\rm tot} \ge \mu_b+\nu_b$}
{break\;}
\uElseIf{$k_a^{\rm tot}$ =  $\mu_a$ {\rm and} $k_b^{\rm tot}$ =  $\mu_b$}{
Append $[D^{i}, D^{i+1}]$ to $\mathcal{S}_{[\mu_a,\mu_b]}$~~~~~\tcp{ assign to the coincidence $[\mu_a,\mu_b]$}

}
\uElseIf{ $k_a^{\rm tot}$ =  $2\nu_a$ {\rm and} $k_b^{\rm tot}$ =  $2\nu_b$}{
\If{ $\left(\theta_a^{i+1}-\theta_a^{i}-\theta_b^{i+1}+\theta_b^{i}\right) \mod 2\pi$ = {\rm 0} \rm{or} $\pi$
}
{Append $[D^{i}, D^{i+1}]$ to $\mathcal{S}_{[2\nu_a,2\nu_b]}$~~~~~\tcp{assign to the coincidence $[2\nu_a,2\nu_b]$}
}
}
\Else{Append $[D^{i}, D^{i+1}]$ to $\mathcal{S}_{[k_a^{\rm tot},k_b^{\rm tot}]}$~~~~~\tcp{ assign to  the coincidence $[k_a^{\rm tot},k_b^{\rm tot}]$}
}
}
}
}
\end{algorithm}

\clearpage

\section{Security Proof of Asynchronous MDI-QKD}

We previously provide\zy{d} a simple and intuitive security proof for asynchronous MDI-QKD~\cite{xie2022breaking} using an entanglement distribution-based argument. In this scheme, Charlie prepares Bell states $\ket{\psi^\pm}_{ab}$ and sends optical mode $a$ to Alice and optical mode $b$ to Bob,
\begin{equation}
	\begin{aligned}
		\label{eq_psi_pm1}
		\ket{\psi^\pm}_{ab}&=\frac{1}{\sqrt{2}}e^{\textbf{i}\varphi}
		\left(\ket{10}^{ij}_{a}\ket{01}^{ij}_{b}
		\pm\ket{01}^{ij}_{a}\ket{10}^{ij}_{b}\right) \\
		&=\frac{1}{\sqrt{2}}\left(\frac{\ket{10}_{a}^{ij}\pm e^{\textbf{i}\varphi}\ket{01}_{a}^{ij}}{\sqrt{2}}\frac{\ket{10}_{b}^{ij}+e^{\textbf{i}\varphi}\ket{01}_{b}^{ij}}{\sqrt{2}}-\frac{\ket{10}_{a}^{ij}\mp e^{\textbf{i}\varphi}\ket{01}_{a}^{ij}}{\sqrt{2}}\frac{\ket{10}_{b}^{ij}-e^{\textbf{i}\varphi}\ket{01}_{b}^{ij}}{\sqrt{2}}\right)\\
		&=\frac{1}{\sqrt{2}}
		\big(\ket{10}_{ab}^i\ket{01}_{ab}^j
		\pm\ket{01}_{ab}^i\ket{10}_{ab}^j
		\big),
	\end{aligned}
\end{equation}
where $\ket{10}^{ij}_{a}$ represents $\ket{1}_{a}^{i}\otimes\ket{0}_{a}^{j}$ and is a tensor product of quantum states between $i$ and $j$ time bins, $\ket{10}_{ab}^i$ represents $\ket{1}_{a}^{i}\otimes\ket{0}_{b}^{i}$ and is a tensor product of quantum states between $a$ and $b$ modes, and $\ket{0}$ and $\ket{1}$ are vacuum and single-photon states, respectively. 
The final result in Eq.~\eqref{eq_psi_pm1}
is obtained via the dualism characterising the entanglement of two identical particles~\cite{bose2013duality}. We notice that $i$ and $j$ can be decoupled as two independent variables,
\lzref{without following a specific relationship such as}
the equalities $i=2d-1$ and $j=2d$ where $d$ is a positive integer.
Therefore, asynchronous MDI-QKD \lzref{can} follow the security proof theory of the original MDI-QKD~\cite{lo2012measurement},
\lzref{i.e., the security analysis based on}
time-reversed entanglement distribution~\cite{xu2020secure}. \lzref{Instead of choosing the $\boldsymbol{Z}$ and $\boldsymbol{X}$ bases to measure the received optical modes $a$ and $b$ as entanglement distribution~\cite{bennett1992quantum}, here} Alice and Bob randomly send optical modes $a$ and $b$ to Charlie.
In other words, Alice randomly sends quantum states $\ket{10}_{a}^{ij}$ and $\ket{01}_{a}^{ij}$ $\Big((\ket{10}_{a}^{ij}+ e^{\textbf{i}\varphi}\ket{01}_{a}^{ij})/\sqrt{2}$ and $(\ket{10}_{a}^{ij}- e^{\textbf{i}\varphi}\ket{01}_{a}^{ij})/\sqrt{2}\Big)$ to Charlie if she chooses the $\boldsymbol{Z}$ ($\boldsymbol{X}$) basis. \lzref{That is to say,} they can independently generate quantum states by using a phase-randomized coherent state instead of a single-photon state between $i$ and $j$ time bins.
Then, one can post-match the successful click events and encode the information in the same way as the original MDI-QKD~\cite{lo2012measurement}, and we call this protocol \emph{asynchronous}.
Note that the same idea of \emph{post-matching} \zy{was independently} proposed in Ref.~\cite{zeng2022mode}, \zy{which gave} a detailed security proof. 

The \lzref{security analysis based on} entanglement distribution~\cite{bennett1992quantum} is easy to understand, but it may not be rigorous enough to use it directly for security proofs.
Here, we provide a \zy{rigorous} 
security proof \lzref{for} asynchronous MDI-QKD utilizing the entanglement swapping argument and its equivalent protocol reduction. We shall start from our virtual protocol 1, where Alice and Bob prepare the non-maximally entangled state. 

\textbf{Virtual protocol 1:} (i) At each time bin $d$, Alice (Bob) prepares an entangled state $\ket{\phi}_{Aa(Bb)}^{d}=\sqrt{t}\ket{+z}_{A(B)}^{d}\ket{1}_{a(b)}^{d}+\sqrt{1-t}\ket{-z}_{A(B)}^{d}\ket{0}_{a(b)}^{d}$, where $\ket{\pm z}_{A (B)}^{d}$ are two eigenvectors of the $\boldsymbol{Z}$ basis denoting the $d$-th qubits and $0<t<1$. Accordingly, we have two eigenvectors of the $\boldsymbol{X}$ basis $\ket{\pm x_{\varphi}}=(\ket{+z}\pm e^{\textbf{i}\varphi}\ket{-z})/\sqrt{2}$ with relative phase $\varphi\in[0,\pi)$.
Let $A^{d}$ and $a^{d}$ represent the $d$-th qubit and the $d$-th optical mode, respectively. Alice and Bob each \lzref{keep} their qubits $A^d$ and $B^d$ in quantum memories and send optical modes $a^d$ and $b^d$ to Charlie via an insecure quantum channel. (ii) Charlie performs single-photon interference measurement with a 50/50 beam splitter and two detectors. When only one detector \lys{clicks}, Charlie obtains a successful click event. He then announces the time bin position \hly{and its corresponding detector}. (iii) Alice and Bob arbitrarily match two click events as a successful asynchronous Bell measurement result and identify \lzref{the corresponding Bell state.} (iv) When the two time bins $i$ and $j$ are selected to be matched (known by Eve), Alice (Bob) performs \lzref{a} controlled-NOT operation \lzref{(controlled by $A^i$ ($B^i$))} on $A^j$ ($B^j$). Then, Alice and Bob \lzref{respectively} measure qubits $A^j$ and $B^j$ in the $\boldsymbol{Z}$ basis, and keep qubits $A^i$ and $B^i$ if their measurement results are both $-1$. \lzref{Finally,} the kept qubits \lzref{in} Alice and Bob, $A^i$ and $B^i$, \lzref{will} establish entanglement. (v) Alice (Bob) randomly chooses the $\boldsymbol{Z}$ and $\boldsymbol{X}$ bases with probabilities $p_{z}$ and $p_{x}=1-p_{z}$ to measure qubits $A^i$ ($B^i$) and obtains the corresponding bit values. \lzref{Bob decides whether he does \lys{bit flip} based on the Bell measurement result.}
(vi) 
Alice and Bob publish the basis information through an authenticated classical channel. 
\lzref{The data in the $\boldsymbol{Z}$ basis and $\boldsymbol{X}$ basis are used to generate raw key bits and estimate the phase error rate, respectively.}
By implementing error correction and privacy amplification, Alice and Bob can distill the final secure key.

\lzref{Because Alice and Bob's operations commute with Charlies', they can} 
perform \hly{the} \lzref{matching} and controlled-NOT operation before Charlie applies single-photon interference measurement, 
\lzref{which will not change}
the security of the protocol. Thus, we have the following virtual protocol 2, whose security is equivalent to virtual protocol 1 from the view of Eve.

\textbf{Virtual protocol 2:} (i) \lzref{Alice (Bob) prepares the entangled state $\ket{\phi}_{Aa(Bb)}^{d}$ and keeps it in quantum memory.}
They \lzref{pre-match the $i$-th and $j$-th time bins}
(known by Eve) and perform a controlled-NOT operation \lzref{(controlled by the $i$-th qubit)} on the $j$-th qubit.
\lzref{Then} Alice and Bob measure the $j$-th qubit in the $\boldsymbol{Z}$ basis and keep the $i$-th qubit if their measurement \lzref{results} are both $-1$. (ii) \lzref{According to the sequence in time bins,} Alice and Bob send optical modes $a$ and $b$ to Charlie via an insecure quantum channel.
(iii) Charlie performs single-photon interference measurement with a \lzref{50/50} beam splitter and two detectors. When only one detector clicks, Charlie obtains a successful click event. He then announces the time bin position \lys{and its corresponding detector}. (iv) Alice and Bob identify which \lzref{specific} Bell state was measured according to the click events at time bins $i$ and $j$. Steps (v) and (vi)
are the same with \lzref{that} in the virtual protocol 1.

Note that the security has no change even if each pairing time bin $i$ and $j$ are 
broadcasted to Alice and Bob \lzref{by Charlie} in step (i).
\lzref{Because} the security of virtual protocol 2 comes from \lzref{that} Alice and Bob randomly choosing (independent of Charlie's operation) the $\boldsymbol{Z}$ and $\boldsymbol{X}$ bases to measure their kept virtual qubits.
\lzref{The pairing time bins $i$ and $j$ are pre-matched in virtual protocol 2. However,}
Alice and Bob can perform the post-matching to increase the valid data (considering the large transmission loss) in Virtual Protocol 1. \lzref{Hence} the efficiency \lzref{in} virtual protocols 1 and 2 are different due to the transmission loss of optical modes. The security does not change since we only consider the successful match\lzref{ing} results, and the other cases will be ruled out. Additionally, there is a similar story in the original MDI-QKD, \lzref{i.e.,} the transmission loss is independent of the security proof. 
\lzref{More intuitively,}
see how entanglement swapping is performed in virtual protocol 2 as below. If the time bins $i$ and $j$ are 
matched, the corresponding joint quantum state of Alice can be written as
\begin{equation}
	\begin{aligned}
		\ket{\phi}_{Aa}^{i}\otimes	\ket{\phi}_{Aa}^{j}
		&=t\ket{+z+z}_{A}^{ij}\ket{11}_{a}^{ij}+(1-t)\ket{-z-z}_{A}^{ij}\ket{00}_{a}^{ij}\\
		&+\sqrt{t(1-t)}\left(\ket{+z-z}_{A}^{ij}\ket{10}_{a}^{ij}+\ket{-z+z}_{A}^{ij}\ket{01}_{a}^{ij}\right),\\
	\end{aligned}
\end{equation}
where quantum state $\ket{+z+z}_{A}^{ij}$ represents $\ket{+z}_{A}^{i}\otimes\ket{+z}_{A}^{j}$.
\lzref{When Alice measures $A_j$ in the $\boldsymbol{Z}$ basis and obtains the outcome $-1$,} he then performs \lzref{a} controlled-NOT operation and keeps the quantum state, 
\lzref{which is normalized as}
\begin{equation}
	\begin{aligned}
		\ket{\Phi}_{Aa}
		=&\frac{1}{\sqrt{2}}e^{\textbf{i}2\varphi}\left(\ket{+z}_{A}^{i}\ket{10}_{a}^{ij}+\ket{-z}_{A}^{i}\ket{01}_{a}^{ij}\right)\\
		=&\frac{1}{\sqrt{2}}\left(\ket{+x_{\varphi}}^{i}_A\frac{\ket{10}^{ij}_a+e^{\textbf{i}\varphi}\ket{01}^{ij}_a}{\sqrt{2}}+\ket{-x_{\varphi}}^{i}_A\frac{\ket{10}^{ij}_a-e^{\textbf{i}\varphi}\ket{01}^{ij}_a}{\sqrt{2}}\right).
	\end{aligned}
\end{equation}
The joint quantum state $\ket{\phi}_{Aa}$ can be regarded as a Bell state since quantum states $\ket{01}^{ij}$ and $\ket{10}^{ij}$ can be denoted as two eigenvectors of the $\boldsymbol{Z}$ basis. Similarly, Bob will have the joint quantum state $\ket{\Phi}_{Bb}=\frac{1}{\sqrt{2}}(\ket{+z}_{B}^{i}\ket{10}_{b}^{ij}+\ket{-z}_{B}^{i}\ket{01}_{b}^{ij})$. Alice and Bob's qubits will \lzref{establish} entanglement after Charlie performs a Bell measurement, that is, two
\lzref{respective} single-photon interferences in time bins $i$ and $j$. This is exactly the entanglement swapping.

We remark that the measurement in step (v) can be applied after step (i) since the measurement operation commutes with all the operations in other steps. Note that for the Bell state $\ket{\Phi}_{Aa}$, Alice sends the optical modes $\ket{10}_{a}^{ij}$ $\left(\ket{01}_{a}^{ij}\right)$ to Charlie if her measured qubit is $\ket{+z}_{A}^{i}$ $\left(\ket{-z}_{A}^{i}\right)$. Similarly, Alice sends the optical modes $\frac{1}{\sqrt{2}}(\ket{10}_{a}^{ij}+ e^{\textbf{i}\varphi}\ket{01}_{a}^{ij}$) $\left(\frac{1}{\sqrt{2}}(\ket{10}_{a}^{ij}- e^{\textbf{i}\varphi}\ket{01}_{a}^{ij})\right)$to Charlie if 
her measured qubit is $\ket{+x_{\varphi}}_{A}^{i}$ $\left(\ket{-x_{\varphi}}_{A}^{i}\right)$. Instead of preparing \lzref{an} entangled state and performing \lzref{a} controlled-NOT operation, one can directly prepare \lzref{and measure} the optical mode. Therefore, virtual protocol 2 is mathematically equivalent to a prepare-and-measure protocol, i.e., virtual protocol 3.

\textbf{Virtual protocol 3:} (i) For each \lzref{matched} two time bins, such as $i$ and $j$, Alice (Bob) randomly prepares optical modes $\ket{10}_{a(b)}^{ij}$ and $\ket{01}_{a(b)}^{ij}$ when she chooses the $\boldsymbol{Z}$ basis with probability $p_{z}$, while Alice (Bob) randomly prepares optical modes $\frac{1}{\sqrt{2}}\left(\ket{10}_{a(b)}^{ij}+ e^{\textbf{i}\varphi}\ket{01}_{a(b)}^{ij}\right)$ and $\frac{1}{\sqrt{2}}\left(\ket{10}_{a(b)}^{ij}- e^{\textbf{i}\varphi}\ket{01}_{a(b)}^{ij}\right)$ when she chooses the $\boldsymbol{X}$ basis with probability $p_{x}$. \lzref{Steps (ii) - (vi) are the same with that in the virtual protocol 2.}


Note that virtual protocol 3 is \lzref{exactly} the original MDI-QKD with a single-photon source if we have $i=2d-1$ and $j=2d$. All optical modes $\ket{10}^{ij}$, $\ket{01}^{ij}$, $\frac{1}{\sqrt{2}}(\ket{10}^{ij}+e^{\textbf{i}\varphi} \ket{01}^{ij})$ and $\frac{1}{\sqrt{2}}(\ket{10}^{ij}-e^{\textbf{i}\varphi} \ket{01}^{ij})$ use a single photon as the information carrier, \lzref{establishing} the joint quantum state between time bins $i$ and $j$. In practice, we usually use a \lzref{weak coherent} laser source to replace the single-photon source
\lzref{in a conventional MDI-QKD setup.}
\lzref{When} the global phase 
$\theta$ is random and unknown to Eve, the phase-randomized coherent state $\ket{e^{\textbf{i}\theta}\sqrt{k}}$ is a mixture of Fock states with mean photon number $k$. 
The user \lzref{then} generate a phase-randomized weak coherent state with a intensity $k_{z}$ in time bin $i$ ($j$) and a vacuum state in time bin $j$ ($i$), i.e., $\ket{e^{\textbf{i}\theta}\sqrt{k_{z}}}^{i}\ket{0}^{j}$ $\left(\ket{0}^{i}\ket{e^{\textbf{i}\theta}\sqrt{k_{z}}}^{j}\right)$. Using the tagging model~\cite{gottesman2004security}, one can securely consider the joint single-photon component, which is equivalent to generating optical modes $\ket{10}_{a}^{ij}$ ($\ket{01}_{a}^{ij}$) with probability $e^{-k_{z}}$. Similarly, the user can generate phase-randomized weak coherent states in time bins $i$ and $j$ with the same intensity $k_{x}$ and relative phase difference $\varphi$ ($\varphi+\pi$), i.e., $\ket{e^{\textbf{i}\theta}\sqrt{k_{x}}}^{i}\ket{e^{\textbf{i}(\theta+\varphi)}\sqrt{k_{x}}}^{j}$ $\left(\ket{e^{\textbf{i}\theta}\sqrt{k_{x}}}^{i}\ket{-e^{\textbf{i}(\theta+\varphi)}\sqrt{k_{x}}}^{j}\right)$. 
\lzref{The joint single-photon component is securely equivalent to optical modes $\frac{1}{\sqrt{2}}(\ket{10}^{ij}+e^{\textbf{i}\varphi} \ket{01}^{ij})$ $\left(\frac{1}{\sqrt{2}}(\ket{10}^{ij}-e^{\textbf{i}\varphi} \ket{01}^{ij})\right)$ with probability $e^{-2k_{x}}$.}


Therefore, we have a virtual protocol 4 with phase-randomized coherent state source, where the joint single-photon component is equivalent to \lzref{that in} virtual protocol 3. 

\textbf{Virtual protocol 4:} (i) For each selected pairing time bins, such as $i$ and $j$, Alice (Bob) randomly prepares phase-randomized coherent states $\ket{e^{\textbf{i}\theta}\sqrt{k_{z}}}^{i}\ket{0}^{j}$ and $\ket{0}^{i}\ket{e^{\textbf{i}\theta}\sqrt{k_{z}}}^{j}$ when she chooses the $\boldsymbol{Z}$ basis with probability $p_{z}$, while she (he) randomly prepares phase-randomized coherent states $\ket{e^{\textbf{i}\theta}\sqrt{k_{x}}}^{i}\ket{e^{\textbf{i}(\theta+\varphi)}\sqrt{k_{x}}}^{j}$ and $\ket{e^{\textbf{i}\theta}\sqrt{k_{x}}}^{i}\ket{-e^{\textbf{i}(\theta+\varphi)}\sqrt{k_{x}}}^{j}$ when she chooses the $\boldsymbol{X}$ basis with probability $p_{x}$. The other steps are the same with steps (ii)-(v) in the virtual protocol 3 except that we need to only consider the contribution of single-photon component. 

There is a tensor product relationship among coherent states in different time bins $\bigotimes_{d=1}^{N}\ket{e^{\textbf{i}\theta_{d}}\sqrt{k_{d}}}^{d}$, and each coherent state has a random but definite phase $\theta_{d}$ and intensity $k_{d}$ from the view of Charlie. Obviously, with the coherent state as a carrier, the sent quantum state is completely decoupled from all time bins in the prepare-and-measure protocol. The intensity and phase of each coherent state are random, and their information is stored in classical memory. This allows us to post-match phase-randomized coherent optical pulses at any two time bins (the information $i$ and $j$ can be known or determined by Charlie) as a joint quantum state $\ket{e^{\textbf{i}\theta_{i}}\sqrt{k_{i}}}^{i}\ket{e^{\textbf{i}\theta_{j}}\sqrt{k_{j}}}^{j}$. Alice and Bob will obtain the key bit and the basis based on the post-matching joint quantum state.
\lzref{Note that the bases and bits are random from the view of Eve.}

\lys{In virtual protocol 4, we pre-determine pairing time bins and perform quantum state preparation 
according to the randomly chosen basis, key bit and intensity. 
When post-matching quantum states in two time bins to acquire the corresponding \emph{random} basis, key bit and intensity, the security of protocol will not change. Because Charlie will not obtain extra \lzref{information from quantum and classical steps}, i.e., \emph{the identical phase-randomized coherent states in quantum communication steps, and the corresponding basis and intensity in classical communication steps}.} For example, 
\lzref{when selecting the $\boldsymbol{Z}$ basis, key bit $0$ and intensity $k_{z}$ for pre-determined pairing time bins $i$ and $j$,}
the quantum state $\ket{e^{\textbf{i}\theta}\sqrt{k_{z}}}^{i}\ket{0}^{j}$ will be \lzref{prepared} and sent to Charlie.
Therein, the intensity and basis will be announced to Charlie in classical communication step.
For post-matching joint quantum state $\ket{e^{\textbf{i}\theta}\sqrt{k_{z}}}^{i'}\ket{0}^{j'}$ at time bins $i'$ and $j'$, one will have the intensity $k_{z}$ and then be assigned \lzref{as} the $\boldsymbol{Z}$ basis and bit value $0$. \lzref{Similarly}, the intensity and basis \lzref{are} announced to Charlie. In the above two cases, 
Eve obtains the same quantum and classical information.
\lzref{Hence, in order}
to obtain as many valid results as possible, we choose to perform a post-matching \lzref{scheme} after Charlie announces all click events in our asynchronous MDI-QKD.

\begin{figure}[t]
\includegraphics[width=0.9\columnwidth]{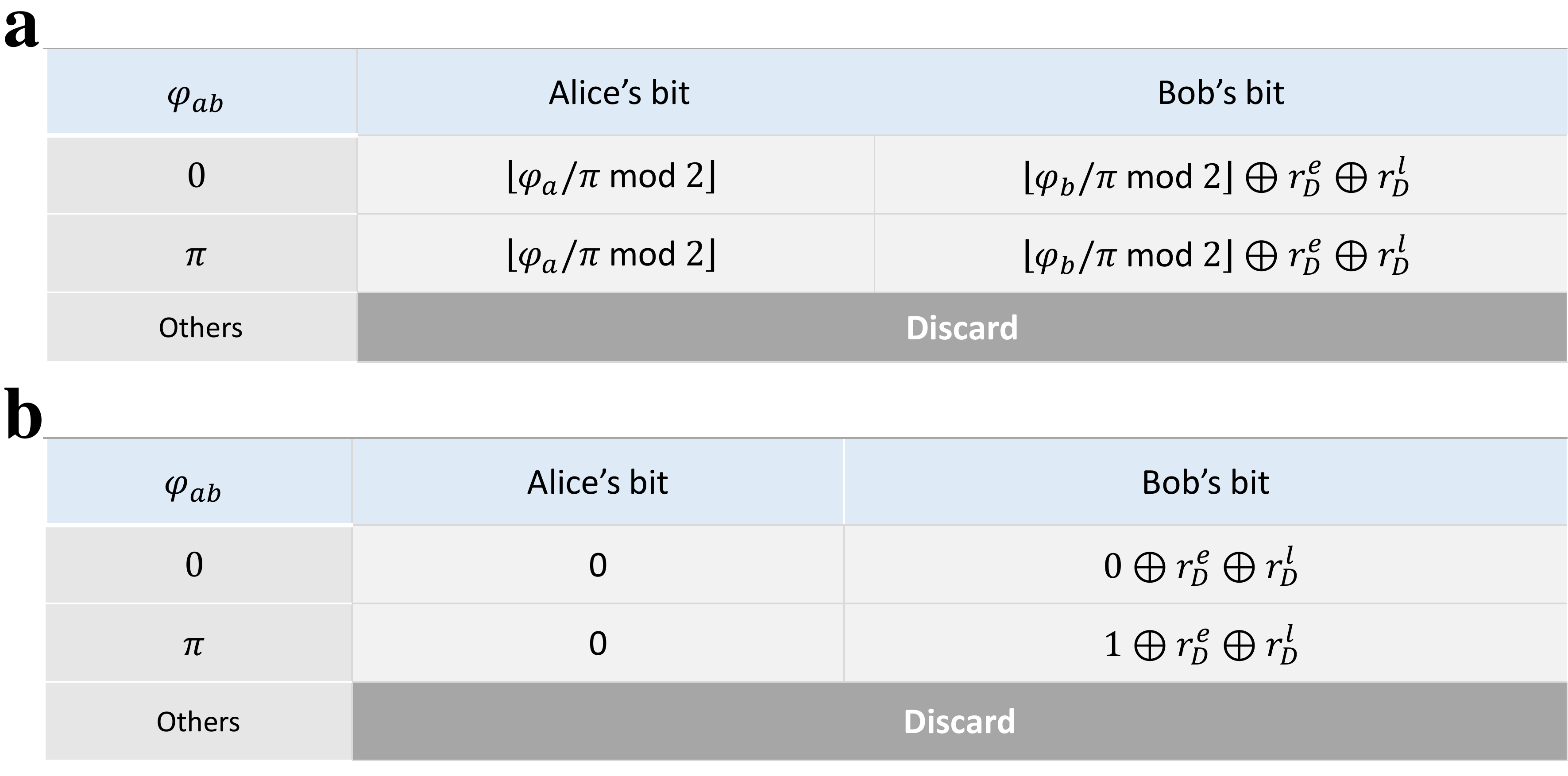}
\caption{\textbf{Two cases of key mapping in the $X$ basis.} \textbf{a}, Corresponding to the virtual protocol 4; We set $r_D$ as bit 0 (1) if detector $D_L$ ($D_R$) clicks. Alice and Bob calculate the relative phase difference $\varphi_{ab}=(\varphi_{a}-\varphi_{b})\mod2\pi$. If $\varphi_{ab} = 0$ or $\pi$, Alice extracts an $\boldsymbol{X}$-basis bit $\left \lfloor\varphi_{a}/\pi\mod2\right \rfloor$, and Bob extracts an $\boldsymbol{X}$-basis bit $\left \lfloor\varphi_{b}/\pi\mod2\right \rfloor \oplus r_D^e\oplus r_D^l$,  where  the superscript $e$~($l$) refers to the early  (late) time bin, respectively. The coincidences with other phase differences are discarded.
\textbf{b}, The actual protocol in the main text; If $\varphi_{ab} = 0$ or $\pi$,  Alice and Bob extract an $\boldsymbol{X}$-basis bit 0. Then, Bob will flip his bit value if $\varphi_{ab} = 0$ and both detectors clicked or $\varphi_{ab} = \pi$ and the same detector clicked twice. The other pairing results are discarded. Obviously, these two schemes (\textbf{a}) and (\textbf{b}) have the identical QBER.
}\label{fig:S1}
\end{figure}

In conclusion, the post-matching single-photon component states \lzref{in} the $\boldsymbol{Z}$ and $\boldsymbol{X}$ bases are randomly generated and independent of Charlie, even if the  post-matching time bins are determined by Charlie. To simplify key mapping of the $X$ basis in our main text, instead of mapping quantum states $\ket{e^{\textbf{i}\theta}\sqrt{k_{x}}}^{i}\ket{e^{\textbf{i}(\theta+\varphi)}\sqrt{k_{x}}}^{j}$ and $\ket{e^{\textbf{i}\theta}\sqrt{k_{x}}}^{i}\ket{-e^{\textbf{i}(\theta+\varphi)}\sqrt{k_{x}}}^{j}$ to $0$ and $1$, respectively, as shown in Fig.~\ref{fig:S1}\textbf{a}, we make both states represent $0$ as shown in Fig.~\ref{fig:S1}\textbf{b}. Note that the QBER of $\boldsymbol{X}$ basis between Fig.~\ref{fig:S1}\textbf{a} and Fig.~\ref{fig:S1}\textbf{b} are identical.
So for, we have proven that asynchronous MDI-QKD with single-photon component is equivalent to Virtual Protocol 1.
\lzref{Furthermore}, to estimate the contribution of a single-photon component, we need to exploit the decoy-state method~\cite{hwang2003quantum, wang2005beating, lo2005decoy}. In the experiment, we can utilize discrete phase randomization to replace continuous phase randomization\cite{cao2015discrete}, such as the number of phase \lzref{slices} $M=16$. 

\section{Secrecy analysis} 
We define $\boldsymbol{E}'$ as all information of Eve learned from raw key $\mathcal{Z}$. 
Alice and Bob can extract a $\varepsilon_{\rm{sec}}$-secret key of length $\ell$ from $\mathcal{Z}$~\cite{tomamichel2012tight},
\begin{equation}
\begin{aligned}
\varepsilon_{\rm{sec}}=2\varepsilon+\frac{1}{2}\sqrt{2^{\ell-H_{\rm{min}}^\varepsilon(\mathcal{Z}|\boldsymbol{E}')}},
\end{aligned}\label{se_eq_1}
\end{equation}
where $H_{\rm{min}}^\varepsilon(\mathcal{Z}|\boldsymbol{E}')$ is the smooth min-entropy, which quantifies the max probability that Eve guesses $\mathcal{Z}$ correctly giving $\boldsymbol{E}'$. According to a chain-rule inequality for smooth entropies, we have 
$ H_{\rm{min}}^\varepsilon(\mathcal{Z}|\boldsymbol{E}')\ge  H_{\rm{min}}^\varepsilon(\mathcal{Z}|\boldsymbol{E})-\lambda_{\rm{EC}}-\log_{2}(2/\varepsilon_{\rm{cor}})$,
where $\boldsymbol{E}$ denotes the information of Eve before error correction, and $\lambda_{\rm{EC}}+\log_2(2/\varepsilon_{\rm{cor}})$ refers to the amount of bit information during the error correction step. The bits of $\mathcal{Z}$ can be distributed among three different strings: $\mathcal{Z}_0$, $\mathcal{Z}_{11}$ and $\mathcal{Z}_{\rm{rest}}$, where $\mathcal{Z}_0$ is the bits where Alice sent a vacuum state, $\mathcal{Z}_{11}$ is the bits where both Alice and Bob sent a single photon and $\mathcal{Z}_{\rm{rest}}$ is the rest of bits. Using a chain-rule for smooth entropies~\cite{Vitanov2013chain}, we have 
\begin{equation}
\begin{aligned}
 H_{\rm{min}}^\varepsilon(\mathcal{Z}|\boldsymbol{E})&\ge  H_{\rm{min}}^{\varepsilon'+2\varepsilon_e+(\hat{\varepsilon}+2\hat{\varepsilon}'+\hat{\varepsilon}'')}(\mathcal{Z}_{0}\mathcal{Z}_{11}\mathcal{Z}_{\rm{rest}}|\boldsymbol{E})\\
 &\ge s_0^z+H_{\rm{min}}^{\varepsilon_e}(\mathcal{Z}_{11}|\mathcal{Z}_{0}\mathcal{Z}_{\rm{rest}}\boldsymbol{E})-2\log_2\frac{2}{\varepsilon'\hat{\varepsilon}},
\end{aligned}\label{se_eq_3}
\end{equation}
where $\varepsilon=\varepsilon'+2\varepsilon_e+(\hat{\varepsilon}+2\hat{\varepsilon}'+\hat{\varepsilon}'')$, and we have used the fact that $H_{\rm{min}}^{\hat{\varepsilon}'}(\mathcal{Z}_{\rm{rest}}|\mathcal{Z}_{0}\boldsymbol{E})\ge0$, $H_{\rm{min}}^{\hat{\varepsilon}''}(\mathcal{Z}_{0}|\boldsymbol{E})\ge H_{\rm{min}}(\mathcal{Z}_{0})=s_0^z$. Here, we consider that vacuum state is uniformly distributed and has no information.

We denote the $\boldsymbol{Z}$ basis as $\ket{10}$, $\ket{01}$, and $\boldsymbol{X}$ as $\frac{1}{\sqrt{2}}(\ket{10}+e^{\textbf{i}\varphi}\ket{10})$, $\frac{1}{\sqrt{2}}(\ket{10}-e^{\textbf{i}\varphi}\ket{10})$.
Taking that Alice and Bob use the bit strings $\mathcal{X}_{11}$ and $\mathcal{X}_{11}'$ of length $s_{11}^z$ to replace the raw key bit string $\mathcal{Z}_{11}$ and   $\mathcal{Z}_{11}'$, we can rewrite the quantity $H_{\rm{min}}^{\varepsilon_e}(\mathcal{Z}_{11}|\mathcal{Z}_{0}\mathcal{Z}_{\rm{rest}}\boldsymbol{E})$ 
using the entropic uncertainty relation~\cite{tomamichel2012tight,curty2014finite}: 
\begin{equation}
\begin{aligned}
H_{\rm{min}}^{\varepsilon_e}(\mathcal{Z}_{11}|\mathcal{Z}_{0}\mathcal{Z}_{\rm{rest}}\boldsymbol{E})&\ge s_{11}^z-H_{\rm{max}}^{\varepsilon_e}(\mathcal{X}_{11}|\mathcal{X}_{11}')\\
&\ge s_{11}^z\left[1-H_2(\phi_{11}^z)\right],
\end{aligned}\label{se_eq_4}
\end{equation}
where $H_2(\phi_{11}^z)$ in the second equation quantifies the required number of bits that Bob uses bit string $\mathcal{X}_{11}'$ to reconstruct $\mathcal{X}_{11}$. 
Combining Eqs.~(\ref{se_eq_1})-(\ref{se_eq_4}), we have
\begin{equation}
\begin{aligned}
 H_{\rm{min}}^{\varepsilon}(\mathcal{Z}|\boldsymbol{E}')&\ge s_0^z+s_{11}^z\left[1-H_2(\phi_{11}^z)\right]-\lambda_{\rm{EC}}\\
 &-\log_2\frac{2}{\varepsilon_{\rm cor}}-\log_2\frac{2}{\varepsilon'\hat{\varepsilon}}-2\log_2\frac{1}{2\varepsilon_{\rm PA}},
\end{aligned}
\end{equation}
and $\varepsilon_{\rm sec}=2(\varepsilon'+2\varepsilon_e+\hat{\varepsilon}+2\hat{\varepsilon}'+\hat{\varepsilon}'')+\varepsilon_{\rm PA}$.
Finally, setting $\hat{\varepsilon}'=\hat{\varepsilon}''=0$, and considering the failure probabilities for estimating the terms of $s_0^z$, $s_{11}^z$, and \ymx{$e_{11}^x$}, $\varepsilon_0$,  $\varepsilon_1$ and $\varepsilon_\beta$ respectively, we have $\varepsilon_{\rm sec}=2(\varepsilon'+2\varepsilon_e+\hat{\varepsilon})+\varepsilon_0+\varepsilon_1+\varepsilon_\beta+\varepsilon_{\rm PA}$. The protocol is $\varepsilon_{
\rm tol}$-secure, and $\varepsilon_{
\rm tol}=\varepsilon_{\rm cor}+\varepsilon_{
\rm sec}$. 
\section{The post-matching principle}

There are many \zy{choices} 
of pairing strategies, but they should meet two \emph{independence} principles \zy{in order to fulfil the requirements by the security proof and by the decoy-state method}.
\zy{First,} the basis information of each pairing is random and independent of Charlie's (Eve's) operation. Then, one can exploit the bit error rate $e_{11}^{x}$ of the $\boldsymbol{X}$ basis to bound the phase error rate $\phi_{11}^{z}$ of the $\boldsymbol{Z}$ basis. 
\zy{Second,} the total intensity information of each \zy{coincidence pair} under the $\boldsymbol{Z}$ ($\boldsymbol{X}$) basis is random and independent of Charlie's (Eve's) operation. Then, one can utilize the decoy-state method for the joint quantum state \zy{and} 
directly estimate the number of single-photon pairs resulting in the $\boldsymbol{Z}$ basis $s_{11}^{z}$, and the bit error rate of the single-photon pair results in the $\boldsymbol{X}$ basis $e_{11}^{x}$.

We argue that the pairing strategy in our previous proposal~\cite{xie2022breaking} does not directly satisfy the second independence principles. In Ref.~\cite{xie2022breaking}, Alice (Bob) first announces whether she (he) sent a decoy intensity $\nu_{a}$ ($\nu_{b}$) optical pulse. Then, they can actively pair two coherent states to obtain the basis information according to the intensity. The gain of a single-photon pair in the coincidence is indirectly estimated via the gain of a single photon in the click events of each time bin. Although these operations can increase the amount of coincidence, it has to add a security assumption that the single-photon distributions in all click events are independent and identically distributed. We remark that the extra security assumption can be removed for the pairing strategy in Ref.~\cite{xie2022breaking} when the secure key is extracted \lzref{according to only one condition, i.e.,}
maximum pairing time interval $T_{c}$ (a large enough $T_{c}$ usually requires phase locking and phase tracking). \zy{We acknowledge that} 
the pairing strategy in Ref.~\cite{zeng2022mode} meets the above two independence principles because pairing the two nearest click events is independent of the optical pulse intensity; thus, the basis and the corresponding total intensity of each pairing are random.

Here, over our previous proposal~\cite{xie2022breaking}, we provide a new pairing strategy that meets the above two independence principles \lzref{and also} removes the extra security assumption. We keep the step that Alice (Bob) first announces whether she (he) sent a decoy intensity $\nu_{a}$ ($\nu_{b}$) optical pulse. For each click event, if Alice announces the decoy intensity $\nu_{a}$, Bob will publish the click event $(\nu_{a}|\mu_{b})$. Similarly, Alice will publish the click event $(\mu_{a}|\nu_{b})$ if Bob announces that the decoy intensity $\nu_{b}$ is prepared. They discard the click events $(\mu_{a}|\nu_{b})$ and $(\nu_{a}|\mu_{b})$ while keeping all other click events. Any two of all reserved click events can be matched as a pair randomly without considering the intensity of each click event. The nearest pairing is usually the optimal scheme, since the average interval of coincidence should 
\lzref{be the shortest}
to decrease the error rate in the $\boldsymbol{X}$ basis.

Let us explain more clearly why the new strategy satisfies the above two independence principles.
Alice and Bob each have three independent and random intensity optical pulses. Thus, there are nine independent and random click events $(\mu_{a}|\mu_{b})$, $(\mu_{a}|\nu_{b})$, $(\mu_{a}|o_{b})$, $(\nu_{a}|\mu_{b})$, $(\nu_{a}|\nu_{b})$, $(\nu_{a}|o_{b})$, $(o_{a}|\mu_{b})$, $(o_{a}|\nu_{b})$ and $(o_{a}|o_{b})$. We emphasize that Charlie has no information about the intensity before he announces the click event. After applying a simple filtering operation by discarding click events $(\mu_{a}|\nu_{b}) $ and $(\nu_{a}|\mu_{b})$, there are seven independent click events, 
where four click events $(\mu_{a}|\mu_{b})$, $(\mu_{a}|o_{b})$, $(o_{a}|\mu_{b})$ and $(o_{a}|o_{b})$ are private. 
The reserved click events 
are pre-determined by Alice and Bob's random numbers \lys{chosen in state preparation}. Therefore, after performing the pairing \lzref{scheme}, the basis and the corresponding total intensity of each pair are determined by previous random numbers and \lzref{will not be} affected by Charlie. For example, two click events $(\mu_{a}|\nu_{b})$ and $(o_{a}|\nu_{b})$ are matched as a pairing result $[\mu_{a},2\nu_b]$, Alice will have the $\boldsymbol{Z}$ basis and $\mu_{a}$ intensity, while Bob obtains the $\boldsymbol{X}$ basis and $2\nu_{b}$ intensity. The $\boldsymbol{X}$ basis and intensity $2\nu_{b}$ obtained by Bob are determined by Bob's previous random numbers in state preparation. Note that after Charlie's announcement and before pairing operation, Charlie will know the decoy intensity in each click event. However, this information will not change the yield and \zy{QBER}
of any photon number component in coincidence because these values are only related to Charlie's announcement.

\begin{figure}[t]
\includegraphics[width=0.75\columnwidth]{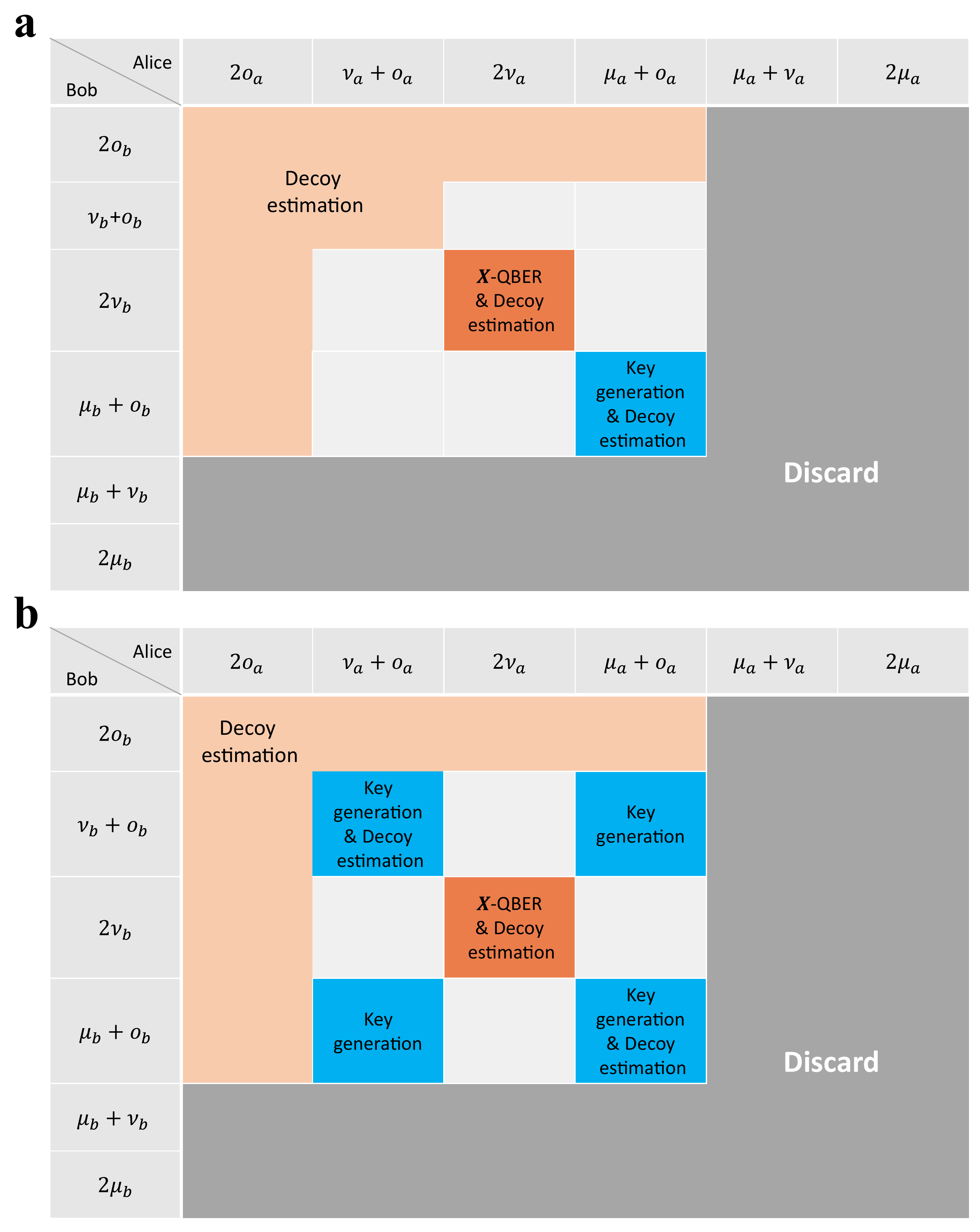}
\caption{ \textbf{Coincidental pairings of asynchronous MDI-QKD.} \textbf{a}, With click filtering, \textbf{b}, Without click filtering. The pairing events  used for key generation, $\boldsymbol{X}$-QBER estimation, decoy estimation and discarded are with blue, orange, pale yellow and gray, respectively. In the $\boldsymbol{X}$ basis, Alice and Bob obtain bit value by calculating the phase difference $\varphi_{ab}=(\varphi_{a}-\varphi_{b})\mod 2 \pi$ and Bob will flip his bit value according to the click detector and $\varphi_{ab}$, as shown in Fig.~\ref{fig:S1}\textbf{b}. Note that $2o_{a(b)}=o_{a(b)}$, $\mu_{a(b)}+o_{a(b)}=\mu_{a(b)}$ and $\nu_{a(b)}+o_{a(b)}=\nu_{a(b)}$ since we have assumed that $o_{a}=o_{b}\equiv0$,  }\label{fig_sheme_despr}
\end{figure}

\section{Basis and key mapping}
In this section, we give \lzref{a} detailed basis and key mapping strategy of the asynchronous MDI-QKD. In Fig.~\ref{fig_sheme_despr}, asynchronous coincidence pairings with and without click filtering are listed. 
\zy{When filtering is applied,  Alice and Bob publish for each photon click whether or not decoy intensity was used in their signal preparation.}
Therefore, detection events $( \nu_a| \nu_b)$, $ (\nu_a|o_b)$ and $ ( o_a|\nu_b)$  are revealed, and detection events $ ( \mu_a|\mu_b)$, $ (\mu_a|o_b)$, $ (o_a|\mu_b)$ or $ (o_a|o_b)$ are unpublished. After post-matching, the raw key is generated from $[\mu_a, \mu_b]$ coincidences (painted in blue). $[2\nu_a, 2\nu_b]$ coincidences are used to calculate the bit error rate in the $X$ basis and make decoy estimation (painted in orange). Whenever one of Alice and Bob pair two click events with one decoy and one signal intensity, or two signal intensities, they discard the pairing event (painted in gray). 
\ymx{
Other cases of pairing events are counted, whose \zy{results} will be used in the decoy state estimation}. In addition, if click filtering is not used, coincidences $[\mu_a, \mu_b]$, $[\mu_a, \nu_b]$, $[\nu_a, \mu_b]$ and $[\nu_a, \nu_b]$ are all kept and can be used for key generation.

\section{Secret key rate calculation} 
\subsection{Key length formula} 
We denote $\underline{x}$ and $\overline{x}$ as the lower and upper bounds of the observed value $x$, respectively, and we denote $x^*$  as the expected value of $x$.
Using the entropic uncertainty relation, the secret key rate $R$ against coherent attacks in the finite-size regime can be written as
\begin{align}
R=&\frac{F}{N}\left\{\underline{s}_{0}^z+\underline{s}_{11}^z\left[1-H_2(\overline{\phi}_{11}^z)\right]-\lambda_{\rm{EC}}-\log_2\frac{2}{\varepsilon_{\rm cor}}-2\log_2\frac{2}{\varepsilon'\hat{\varepsilon}}-2\log_2\frac{1}{2\varepsilon_{\rm PA}}\right\},\label{eq_keyrate}
\end{align}
where $N$ is the data size;  $s^z_{0}$ and $s^z_{11}$  are the number of vacuum components and single-photon pair components in the $\boldsymbol{Z}$ basis, respectively; $\phi_{11}^z$ is single-photon pair phase error rate. \ymx{$\lambda_{\rm{EC}}$ is the amount of information consumed in error correction}; $H_2(x)=-x\log_2x-(1-x)\log_2(1-x)$ is the binary Shannon entropy function. $\varepsilon_{\rm cor}$, $\varepsilon_{\rm PA}$, $\varepsilon'$ and $\hat{\varepsilon}$ are security coefficients regarding the secrecy and correctness.

Let us first introduce the formulas used to calculate the statistical fluctuation. The Chernoff bound is utilized in calculating the finte size effect~\cite{yin2020tight}. 
Given a expected value $x^*$ and specified security parameter $\epsilon$ (failure probability), its upper bound and lower bound of the observed value can be given by \begin{equation}
\begin{aligned} 
	\overline{x}&=x^{*}+\frac{\beta}{2}+\sqrt{2\beta x^{*}+\frac{\beta^{2}}{4}}, \nonumber
\end{aligned}
\end{equation} 
and 
\begin{equation}
\begin{aligned} 
	\underline{x}&=x^{*}-\sqrt{2\beta x^{*}}, \nonumber
\end{aligned}
\end{equation}
where $\beta=\ln{\epsilon^{-1}}$. Similarlity, using the variant of Chernoff bound~\cite{yin2020tight} the upper bound and lower bound of the expected value $x^*$ can also be calculated with a given observed value $x$ and failure probablity $\epsilon$:
\begin{equation}
\begin{aligned}\label{varchernoff1}	\overline{x}^{*}&=x+\beta+\sqrt{2\beta x+\beta^{2}}, \nonumber
\end{aligned}
\end{equation} 
and
\begin{equation}
\begin{aligned}\label{varchernoff2}
	\underline{x}^{*}&=\max\left\{x-\frac{\beta}{2}-\sqrt{2\beta x+\frac{\beta^{2}}{4}},~0\right\}. \nonumber
\end{aligned}
\end{equation}
The random sampling theorem can be applied to give a upper bound of the phase error rate in the Z basis $\phi_{11}^z$, which is given by~\cite{yin2020tight} 
\begin{equation}
\begin{aligned}
	\overline{\chi}\leq&\lambda	+\gamma^{U} (n,k,\lambda,\epsilon), \nonumber
\end{aligned}\label{Randomswr}
\end{equation}
where
\begin{equation}
\gamma^{U}(n,k,\lambda,\epsilon)=\frac{\frac{(1-2\lambda)AG}{n+k}+
	\sqrt{\frac{A^2G^2}{(n+k)^2}+4\lambda(1-\lambda)G}}{2+2\frac{A^2G}{(n+k)^2}}, \nonumber
\end{equation}
with $A=\max\{n,k\}$ and $G=\frac{n+k}{nk}\ln{\frac{n+k}{2\pi nk\lambda(1-\lambda)\epsilon^{2}}}$.

\subsection{Decoy-state estimation}
\subsubsection{Click filtering used}
 As shown in Fig.~\ref{fig_sheme_despr}\textbf{a}, when the click filtering is used, because the information of decoy time bins are announced, only $[\mu_a,\mu_b]$ can be used to generate keys. 
The amount of information consumed in error correction can be written as
\begin{equation}
\begin{aligned}
    \lambda_{\rm{EC}}&=n_{[\mu_a,\mu_b]}fH_2\left(\frac{m_{[\mu_a,\mu_b]}}{n_{[\mu_a,\mu_b]}}\right).
\end{aligned}
\end{equation}

With the help of tagged model,
the final secure secret keys can be distilled from single-photon pair component. It's necessary to first estimate the single-photon pair component.
Using the decoy state method~\cite{hwang2003quantum,wang2005beating,lo2005decoy} and the click filtering, the lower bound of the number of single-photon pair in the $\boldsymbol{Z}$ basis can be given by~\cite{lim2014concise,yin2020tight}
\begin{equation}
\begin{aligned}
\underline{s}_{11}^{z*}&\geq	 \frac{e^{-\mu_a-\mu_b} p_{[\mu_a,\mu_b]}}{\nu_a\nu_b(\mu'-\nu')} \\
&\times\left[\mu_a\mu_b\mu' \left(e^{\nu_a+\nu_b }\frac{\underline{n}_{[\nu_a ,\nu_b] }^{*}}{p_{[\nu_a,\nu_b]}}-e^{\nu_b}\frac{\overline{n}_{[o_a,\nu_b] }^*}{p_{[o_a,\nu_b]}}- 
	e^{\nu_a }\frac{\overline{n}_{[\nu_a ,o_b] }^{*}}{p_{[\nu_a,o_b]}}
	+\frac{\underline{n}_{[o_a,o_b] }^{*}}{p_{[o_a,o_b]}} \right)
\right.\\ 
	&	-\left.\nu_a\nu_b\nu' \left(e^{\mu_a+\mu_b}\frac{\overline{n}_{[\mu_a,\mu_b] }^{*}}{p_{[\mu_a,\mu_b]}}-e^{\mu_b }\frac{\underline{n}_{[o_a,\mu_b] }^*}{p_{[o_a,\mu_b]}}- 
	e^{\mu_a}\frac{\underline{n}_{[\mu_a, o_b] }^{*}}{p_{[\mu_a,o_b]}}
	+ \frac{\underline{n}_{[o_a,o_b] }^{*}}{p_{[o_a,o_b]}} \right)	\right].\label{eq_decoy_Y11}
\end{aligned} 
\end{equation}
Here, 
\begin{equation}
   \begin{cases}
   \mu'=\mu_a,\quad \nu'=\nu_a& \mbox{if} \quad \frac{\mu_a}{\mu_b}
   \le \frac{\nu_a}{\nu_b}, \\
   \mu'=\mu_b, \quad\nu'=\nu_b&\mbox{if}  \quad \frac{\mu_a}{\mu_b}
   > \frac{\nu_a}{\nu_b}.\label{eq_decoy_Y11_sup}
   \end{cases}
  \end{equation}
  and 
\begin{equation}
\begin{aligned}
 p_{[k_a^{\rm tot},k_b^{\rm tot}]}=  \sum\limits_{k_a^e+k_a^l= k_a^{\rm tot}}\sum\limits_{k_b^e+k_b^l=k_b^{\rm tot}}  
  \frac{p_{k_a^e}p_{k_b^e}}{p_s} \frac{p_{k_a^l}p_{k_b^l}}{p_s},
\end{aligned} 
\end{equation}
apart from $p_{[2\nu_a,2\nu_b]}$ because of the phase matching condition in the $\boldsymbol{X}$ basis, which is
\begin{equation}
	\begin{aligned}
	p_{[2\nu_a,2\nu_b]}=\frac{2}{M}\frac{p_{\nu_a}p_{\nu_b}}{p_{s}} \frac{p_{\nu_a}p_{\nu_b}}{p_{s}}.
	\end{aligned} 
\end{equation} 
When using the matching method without click filtering, $p_s=1$; when using the matching method with click filtering, $p_s=1-p_{\mu_a}p_{\nu_b}-p_{\nu_a}p_{\mu_b}$. 
Eqs.~\eqref{eq_decoy_Y11}~-~\eqref{eq_decoy_Y11_sup} are also applicable in asymmetric channels~\cite{wang2019asymmetric}. \ymx{In the symmetric case, $\mu_a=\mu_b$, $\nu_a=\nu_b$ and $o_a=o_b$. Thus we use the total observed values $n_{[\mu_a ,o]}+n_{[o,\mu_b]}$ and $n_{[\nu_a ,o]}+n_{[o,\nu_b]}$ to calculate expected values in Eq.~\eqref{eq_decoy_Y11} to improve SKR due to $p_{[\mu_a,o_b]}=p_{[o_a,\mu_b]}$ and $p_{[\nu_a,o_b]}=p_{[o_a,\nu_b]}$.} The total number of vacuum components in the $\boldsymbol{Z}$  basis  is
\begin{equation}
\begin{aligned}
    \underline{s}_{0}^{z*}& =  
\frac{ e^{-\mu_a}p_{[\mu_a,\mu_b]}}{p_{[o_a,\mu_b]}}\underline{n}_{[o_a,\mu_b]}^{*}.
\end{aligned}
\end{equation}

The upper bound of single-photon pair errors of the $\boldsymbol{X}$ basis is 
\begin{align}
\overline{t}_{11}^{x}\le m_{[2\nu_a,2\nu_b]}-\underline{m}_{[2\nu_a,2\nu_b]}^{0},\label{phi_11z_start}
\end{align}
where \ymx{$m_{[2\nu_a,2\nu_b]}$ } is the observed error bit number in the $\boldsymbol{X}$ basis. The observed value \ymx{$\underline{m}_{[2\nu_a,2\nu_b]}^{0}$} is the lower bound of error bit number in the $\boldsymbol{X}$ basis given that at least one of Alice and Bob sends vacuum component, which corresponding to expected value can be given by
\begin{equation}
\begin{aligned}\label{mo}
\underline{m}_{[2\nu_a,2\nu_b]}^{0*}=e^{-2\nu_a}\frac{p_{[2\nu_a,2\nu_b]}}{2p_{[o_a,2\nu_b]}}\underline{n}_{[o_a,2\nu_b]}^{*}+e^{-2\nu_b}\frac{p_{[2\nu_a,2\nu_b]}}{2p_{[2\nu_a,o_b]}}\underline{n}_{[2\nu_a,o_b]}^{*}-e^{-2\nu_a-2\nu_b} \frac{p_{[2\nu_a,2\nu_b]}}{2p_{[o_a,o_b]}}\overline{n}_{[o_a,o_b]}^{*}.
\end{aligned}
\end{equation}
\ymx{Similarly, we use the observed value ${n}_{[o_a,2\nu_b]}+{n}_{[2\nu_a,o_b]}$ to calculate the expected value in Eq.~\eqref{mo} in symmetric channels.} Therefore, the upper bound of the bit error rate of single-photon pair in the $\boldsymbol{X}$ basis can be calculated: $\overline{e}_{11}^x =\frac{\overline{t}_{11}^x}{\underline{s}_{11}^{x}}$, where the lower bound of the number of single-photon pair in the  $\boldsymbol{X}$ basis is
 \begin{equation}
 \begin{aligned}
\underline{s}_{11}^{x*}&\geq	 \frac{e^{-2\nu_a-2\nu_b} 4p_{[2\nu_a,2\nu_b]}}{\mu_a\mu_b(\mu'-\nu')} \\
&\times\left[\mu_a\mu_b\mu' \left(e^{\nu_a+\nu_b }\frac{\underline{n}_{[\nu_a ,\nu_b] }^{*}}{p_{[\nu_a,\nu_b]}}-e^{\nu_b}\frac{\overline{n}_{[o_a,\nu_b] }^*}{p_{[o_a,\nu_b]}}- 
	e^{\nu_a }\frac{\overline{n}_{[\nu_a ,o_b] }^{*}}{p_{[\nu_a,o_b]}}
	+\frac{\underline{n}_{[o_a,o_b] }^{*}}{p_{[o_a,o_b]}} \right)
\right.\\  
	&	-\left.\nu_a\nu_b\nu' \left(e^{\mu_a+\mu_b}\frac{\overline{n}_{[\mu_a,\mu_b] }^{*}}{p_{[\mu_a,\mu_b]}}-e^{\mu_b }\frac{\underline{n}_{[o_a,\mu_b] }^*}{p_{[o_a,\mu_b]}}- 
	e^{\mu_a}\frac{\underline{n}_{[\mu_a, o_b] }^{*}}{p_{[\mu_a,o_b]}}
	+ \frac{\underline{n}_{[o_a,o_b] }^{*}}{p_{[o_a,o_b]}} \right)	\right],
\end{aligned}
\end{equation}
Using the random sampling without replacement theorem, with a failure probability $\varepsilon_e$, we have the upper bound of single-photon pair phase error rate in the $\boldsymbol{Z}$ basis:
\begin{equation}
	\begin{aligned}
		\overline{\phi}_{11}^{z}\leq& \overline{e}_{11}^x	+\gamma \left(\underline{s}_{11}^z,\underline{s}_{11}^x,\overline{e}_{11}^x,\varepsilon_e\right), \label{phi_11z_stop}
	\end{aligned} 
\end{equation}
where $ \gamma^{U}(n,k,\lambda,\epsilon_e) $ function quantifies the finite-size effect in random sampling without replacement.

\subsubsection{Click filtering unused}
When the click filtering is not utilized as shown in Fig.~\ref{fig_sheme_despr}\textbf{b}, there are four pairing results $[\mu_a,\mu_b]$, $[\mu_a,\nu_b]$, $[\nu_a,\mu_b]$ and $[\nu_a,\nu_b]$ that can be used to generate keys. Therefore, the calculation of single-photon pair component $\underline{s}_{11}^{z*}$ and vacuum contribution $\underline{s}_{0}^{z*}$ is different. The amount of information consumed in error correction is  
\begin{equation}
\begin{aligned}
    \lambda_{\rm{EC}}&=\sum_{k_a\in\{\mu_a,\nu_a\}}\sum_{k_b\in\{\mu_b,\nu_b\}} n_{[k_a,k_b]}fH_2\left(\frac{m_{[k_a,k_b]}}{n_{[k_a,k_b]}}\right),
\end{aligned}
\end{equation}
where $n_{[k_a,k_b]}$ is the total number for set  $\mathcal{S}_{[k_a,k_b]}$, and $m_{[k_a,k_b]}$ is the number of errors for set  $\mathcal{S}_{[k_a,k_b]}$. 
The lower bound of the $s_{11}^{z*}$ and $s_{0}^{z*}$ can be given by  
\begin{equation}
\begin{aligned}
    \underline{s}_{11}^{z*}&= \left(\mu_a\mu_be^{-\mu_a-\mu_b}p_{[\mu_a,\mu_b]}  +\nu_a\nu_be^{-\nu_a-\nu_b}p_{[\nu_a,\nu_b]} + \mu_a\nu_be^{-\mu_a-\nu_b}p_{[\mu_a,\nu_b]}  +\nu_a\mu_be^{-\nu_a-\mu_b}p_{[\nu_a,\mu_b]}\right)      \\
    &\times\frac{1}{\mu_a\mu_b\nu_a\nu_b(\mu'-\nu')}  \left[\mu_a\mu_b\mu' \left(e^{\nu_a+\nu_b }\frac{\underline{n}_{[\nu_a ,\nu_b] }^{*}}{p_{[\nu_a,\nu_b]}}-e^{\nu_b}\frac{\overline{n}_{[o_a,\nu_b] }^*}{p_{[o_a,\nu_b]}}- 
	e^{\nu_a }\frac{\overline{n}_{[\nu_a ,o_b] }^{*}}{p_{[\nu_a,o_b]}}
	+\frac{\underline{n}_{[o_a,o_b] }^{*}}{p_{[o_a,o_b]}} \right)
\right.\\ \nonumber
	&	-\left.\nu_a\nu_b\nu' \left(e^{\mu_a+\mu_b}\frac{\overline{n}_{[\mu_a,\mu_b] }^{*}}{p_{[\mu_a,\mu_b]}}-e^{\mu_b }\frac{\underline{n}_{[o_a,\mu_b] }^*}{p_{[o_a,\mu_b]}}- 
	e^{\mu_a}\frac{\underline{n}_{[\mu_a, o_b] }^{*}}{p_{[\mu_a,o_b]}}
	+ \frac{\underline{n}_{[o_a,o_b] }^{*}}{p_{[o_a,o_b]}} \right)	\right]\\
\end{aligned}
\end{equation}
and
  \begin{equation}
\begin{aligned}
    \underline{s}_{0}^{z*}& = \left(e^{-\mu_a} \frac{p_{[\mu_a,\mu_b]}}{p_{[o_a,\mu_b]}}  +e^{-\nu_a}   \frac{p_{[\nu_a,\mu_b]}}{p_{[o_a,\mu_b]}} \right) \underline{n}_{[o_a,\mu_b]}^{*} +\left( e^{-\mu_a}\frac{p_{[\mu_a,\nu_b]}}{p_{[o_a,\nu_b]}}  +e^{-\nu_a}  \frac{p_{[\nu_a,\nu_b]}}{p_{[o_a,\nu_b]}}  \right) \underline{n}_{[o_a,\nu_b]}^{*} ,
\end{aligned}
\end{equation}
respectively.

The calculation of phase error rate $\overline{\phi}_{11}^{z}$ can directly utilize Eqs.~\eqref{phi_11z_start}~-~\eqref{phi_11z_stop}. 

\subsection{Simulation formulas}
When Alice and Bob send intensities $k_a$ and $k_b$ with phase difference $\theta$, the gain corresponding to only detector $L$ and $R$ click are
\begin{equation}
\begin{aligned}
q_{(k_a|k_b)}^{\theta, L}& =  y_{(k_a|k_b)}^Re^{\eta_d
^R\sqrt{\eta_a k_a\eta_bk_b}\cos{\theta}} \left[1-y_{(k_a|k_b)}^Le^{-\eta_d
^L\sqrt{\eta_a k_a\eta_bk_b}\cos{\theta}}\right],
\end{aligned} 
\end{equation} 
and
\begin{equation}
\begin{aligned}
q_{(k_a|k_b)}^{\theta, R}& = y_{(k_a|k_b)}^Le^{-\eta_d
^L\sqrt{\eta_a k_a\eta_bk_b}\cos{\theta}} \left[1-y_{(k_a|k_b)}^Re^{\eta_d
^R\sqrt{\eta_a k_a\eta_bk_b}\cos{\theta}}\right],
\end{aligned} 
\end{equation}
respectively,  where $\eta_d^L~(\eta_d^R)$ and $p_d^L~(p_d^R)$ are the detection efficiency and the dark count rate of the detector $D_L~(D_R)$;  $y_{(k_a|k_b)}^{L(R)}=\left(1-p_d^{L(R)}\right)e^{-\frac{\eta_d
^{L(R)}\left(\eta_a k_a+\eta_b k_b\right)}{2}}$, $\eta_{a}=10^{-\frac{\alpha l_a}{10}}$ and $\eta_{b}=10^{-\frac{\alpha l_b}{10}}$. By integrated the $\theta$ from 0 to $2\pi$, the overall gain when Alice and Bob send intensities $k_a$ and $k_b$ respectively can be expressed as 
\begin{equation}
\begin{aligned}
q_{(k_a|k_b)}&=\frac{1}{2\pi} \int_0^{2\pi}
\left(q_{(k_a|k_b)}^{\theta, R}+q_{(k_a|k_b)}^{\theta, L}\right) d\theta \\
&=y_{(k_a|k_b)}^LI_0\left(\eta_d^L\sqrt{\eta_a k_a\eta_bk_b}\right)+y_{(k_a|k_b)}^R I_0\left(\eta_d
^R\sqrt{\eta_a k_a\eta_bk_b}\right) -2y_{(k_a|k_b)}^Ly_{(k_a|k_b)}^RI_0\left[(\eta_d
^L-\eta_d
^R)\sqrt{\eta_a k_a\eta_bk_b}\right],
\end{aligned} 
\end{equation} 
where $I_0(x)$ refers to the zero-order modified Bessel function of the first kind.  

Given a time bin with a click event, the probability of at least one click event 
\lzv{following the first-arrived click} within the time interval $T_{c}$ is
\begin{equation}
q_{T_{c}}=1-(1-q_{\rm tot})^{N_{T_{c}}},
\end{equation}
where $q_{\rm tot}$ is the probability of having a click event and $N_{T_{c}}=FT_{c}$ is the number of time bins within time interval $T_{c}$. Therefore, it takes an average of $1+1/q_{T_{c}}$ click events to form a valid pairing. Therefore, the total number of valid successful pairing results is
\begin{equation}
n_{\rm tot}= \frac{Nq_{\rm tot}}{1+1/q_{T_{c}}}.
\end{equation}
Besides, the average of paring interval can be given by
\begin{equation}
\begin{aligned}\label{T_mean}
T_{\rm mean}&=\frac{\sum_{i=0}^{N_{T_c}-1} (i+1)\left(1-q_{\rm tot}\right)^iq_{\rm tot}}{F\sum_{i=0}^{N_{T_c}-1} \left(1-q_{\rm tot}\right)^iq_{\rm tot}}\\
&=\frac{1-N_{T_{c}}q_{\rm tot}(1/q_{T_{c}}-1)}{Fq_{\rm tot}}.
\end{aligned} 
\end{equation}
When using the matching method without click filtering, $q_{\rm tot}=\sum\limits_{k_a,k_b}p_{k_a}p_{k_b}q_{(k_a|k_b)}$; when using the matching method with click filtering, $q_{\rm tot}=\sum\limits_{k_a,k_b}p_{k_a}p_{k_b}q_{(k_a|k_b)}-p_{\mu_a}p_{\nu_b}q_{(\mu_a|\nu_b)}-p_{\nu_a}p_{\mu_b}q_{(\nu_a|\mu_b)}$. 

The total number of set $\mathcal{S}_ {[k_a^{\rm tot}, k_b^{\rm tot}]}$ (except set $\mathcal{S}_{[2\nu_a,2\nu_b]}$ ) is 
 \begin{equation}
\begin{aligned}
 n_{[k_a^{\rm tot}, k_b^{\rm tot}]} =n_{\rm tot}  \sum\limits_{k_a^e+k_a^l= k_a^{\rm tot}}\sum\limits_{k_b^e+k_b^l=k_b^{\rm tot}}  
 \left(\frac{p_{k_a^e}p_{k_b^e}q_{(k_a^e|k_b^e)}}{q_{\rm tot}}\frac{p_{k_a^l}p_{k_b^l} q_{(k_a^l|k_b^l)}}{q_{\rm tot}}\right).
\end{aligned} 
\end{equation}
The total number of set $\mathcal{S}_{[2\nu_a, 2\nu_b]}$  is
 \begin{equation}
\begin{aligned}
 n_{[2\nu_a,2\nu_b]}=   \frac{n_{\rm tot}}{M\pi} \int_0^{2\pi}  \left(\frac{p_{\nu_a}p_{\nu_b}q_{(\nu_a|\nu_b)}^{\theta}}{q_{\rm tot}}\frac{p_{\nu_a}p_{\nu_b}q_{(\nu_a|\nu_b)}^{\theta}}{q_{\rm tot}}\right) d\theta.
\end{aligned} 
\end{equation}
Note that although the phase is discretely modulated, the phase arriving at the Charlie station is continuous and random due to phase drift in the fiber. The total number of errors in the $\boldsymbol{X}$ basis can be written as
 \begin{equation}
\begin{aligned}
m_{[2\nu_a,2\nu_b]}=&\frac{n_{\rm tot}}{M\pi} \int_0^{2\pi}   \Bigg\{  (1-E_{\rm{\rm HOM}})  \frac{p_{\nu_a}^2p_{\nu_b}^2\left[q_{(\nu_a|\nu_b)}^{\theta,L}q_{(\nu_a|\nu_b)}^{\theta+\delta ,R}+q_{(\nu_a|\nu_b)}^{\theta,R}q_{(\nu_a|\nu_b)}^{\theta+\delta ,L}\right]}{q_{\rm tot}^2}\\
&+E_{\rm{\rm HOM}} \frac{p_{\nu_a}^2p_{\nu_b}^2 \left[q_{(\nu_a|\nu_b)}^{\theta,L}q_{(\nu_a|\nu_b)}^{\theta+\delta,L}+q_{(\nu_a|\nu_b)}^{\theta,R}q_{(\nu_a|\nu_b)}^{\theta+\delta ,R}\right]}{{q_{\rm tot}^2}}\Bigg\} d\theta,
\end{aligned} 
\end{equation}
where $E_{\rm{HOM}}$ is the interference misalignment error rate,  and $\delta = T_{\rm mean}(2\pi\Delta \lzv{f}+ \omega_{\rm fib})$ is the phase misalignment resulted from fiber phase drift rate $\omega_{\rm fib}$ and laser frequency difference $\Delta \lzv{f}$. \hly{Compared with Eq. (11) in the main text, one can assume that $\omega_{\rm fib}\approx\sigma$ in our simulation due to very small frequency difference $\Delta\nu$.}

\begin{figure}[ht]
\includegraphics[scale=1]{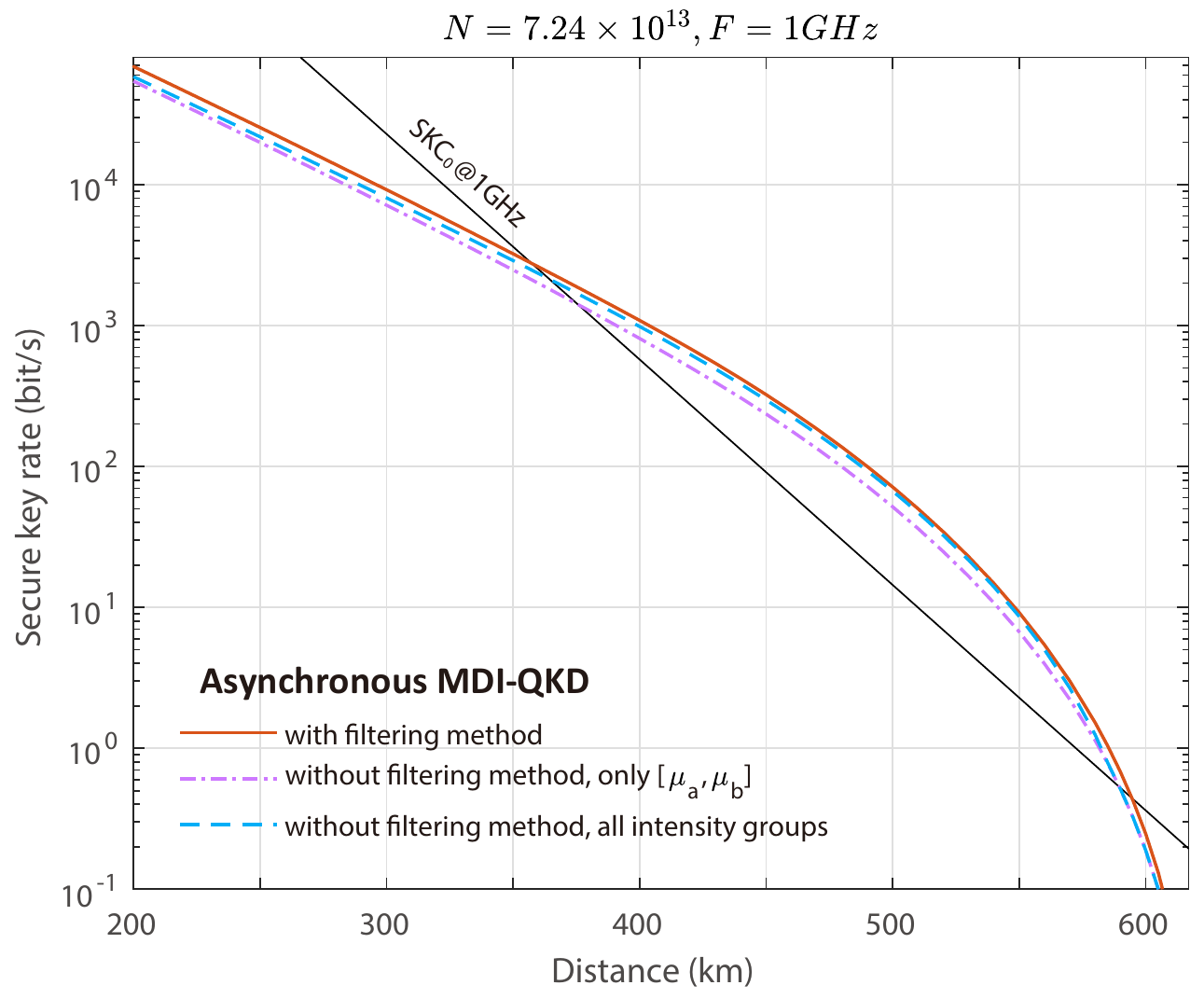}
\caption{\textbf{Comparison of the secret key rates of asynchronous MDI-QKD with click filtering and without click filtering.} The data size is  $N=7.24\times 10^{13}$. The system clock frequency is $F=1$ GHz.}\label{fig_tpnpp}
\end{figure}

We simulate the \zy{SKRs} 
of asynchronous MDI-QKD with \zy{and without} click filtering 
using $[\mu_a,\mu_b]$ for key generation, and without click filtering using all intensity groups for key generation, which is shown in Fig.~\ref{fig_tpnpp}. In this work, we set failure parameters $\varepsilon_{\rm{cor}}$,
$\varepsilon'$, $\varepsilon_e$, $\hat{\varepsilon}$, $\varepsilon_{\beta}$, and $\varepsilon_{\rm PA}$ to be the same
value: $\epsilon$.  We have $
\varepsilon_0+\varepsilon_1=12\epsilon$ because we use the
Chernoff bound 12 times to estimate $s_{0}^z$,
$s_{11}^z$, and $e_{11}^x$. We set $\epsilon=10^{-10}$,  and then $\varepsilon_{\rm{tol}}=\varepsilon_{\rm{cor}}+\varepsilon_{\rm{sec}}=2.3
\times 10^{-9}$. The total number of transmitted quantum signals is set to $7.24\times10^{13}$ (this data sizes is used at 508 km in our experiment). The overall transmittance from Alice (Bob) to Charlie is $\eta_{a}=10^{-(0.16l_a+1.50)/10}$ ($\eta \ymx{_b}=10^{-(0.16l_b+1.50)/10}$), where $l_a$ ($l_b$) is the distance between Alice (Bob) and Charlie, $1.50 $ dB is the insert loss on Charlie's side. $\eta_d^L = 78.1\%$~($\eta_d^R = 77 \%$) and $p_d^L=3.03\times 10^{-9}$ ($p_d^R=3.81\times 10^{-9}$) are the detection efficiency and dark count rate of detector $D_L$ ($D_R$), respectively. The interference misalignment error rate $E_{\rm HOM}=0.04$, fiber phase drift rate $\omega_{\rm fib}=5900$ rad s$^{-1}$, and laser frequency difference $\Delta \lzv{f} =10$ Hz.  The intensities and probabilities were optimized to maximize the key rate per pulse. Simulations show that the protocol with click filtering has a slightly higher key rate than the protocol without click filtering.



\lzv{
The actual scaling depends on the detection probability and the pairing length. According to the Eqs.~(S23) and (S24) in Supplemental Material VI C, the valid successful coincidence probability is
\begin{equation}
\begin{aligned}
P(K)=&\frac{n_{\rm tot}}{N}= \frac{q_{\rm tot}}{1+\frac{1}{1-(1-q_{\rm tot})^{N_{T_{c}}}}}\\
\approx & \frac{\bar{\mu}\eta_d\sqrt{\eta}}{1+\frac{1}{1-(1-\bar{\mu}\eta_d\sqrt{\eta})^{FT_c }}} \\
\propto &\frac{\sqrt{\eta}}{1+1/[1-(1-\sqrt{\eta})^{F  T_{c} }]}
\end{aligned}
\end{equation}
where $q_{\rm tot}\approx \bar{\mu}\eta_d\sqrt{\eta}$ is the detection probability (\lzv{or click probability}) \lzv{for each detection time window};  $\bar{\mu}$ is the total mean photon number of Alice and Bob; $\eta_d$ is the detection efficiency \lzv{of single photon detector}; $N_{T_{c}}=FT_{c}$ is the number of \lzv{detection time windows} within time interval $T_{c}$; $F$ is the system clock frequency. \lzv{Since $FT_c \geq 1$ and $\sqrt{\eta} \le 1-(1-\sqrt{\eta})^{F  T_{c} } < 1$}, We can see that $P(K)\propto \eta$ when $FT_c =1$, and $P(K)\propto \sqrt{\eta}$ when $FT_c \rightarrow\infty$. \lzv{Especially, when there is at least two clicks on average within $Tc$, i.e.,} $F T_c \times(\bar{\mu}\eta_d\sqrt{\eta} ) ~\lzv{\geq}~ 2$, 
we have $(1-\bar{\mu}\eta_d\sqrt{\eta})^{FT_c } ~\lzv{\le}~ 1/e^2$ and \lzv{then $ \frac{(e^2-1)\bar{\mu}\eta_d}{2e^2-1}\sqrt{\eta} \le P(K) < \frac{\bar{\mu}\eta_d}{2}\sqrt{\eta} $,} thus $P(K) \propto\sqrt{\eta}$. In our experiment $F=1$ GHz, $\eta_d = 78\%$,  $\alpha=0.16$ dB/km, we can obtain that $P(K)\approx 0.068\sqrt{\eta}$, $P(K)\approx 0.058\sqrt{\eta}$, $P(K)\approx  0.051\sqrt{\eta}$, and $P(K)\approx 0.072\sqrt{\eta}$ ($T_c=5\mu s$, $20\mu s$, $60\mu s$ and $200\mu s$, $\bar{\mu} =0.18, 0.15, 0.14$, and $0.20$) for 201.86, 306.31, 413.73 and 508.16 km, respectively. \lzv{All the results} scale proportionally to $\sqrt{\eta}$.} 
 
\section{Laser source characterization} 

\begin{figure}[ht]
\includegraphics[width=1\columnwidth]{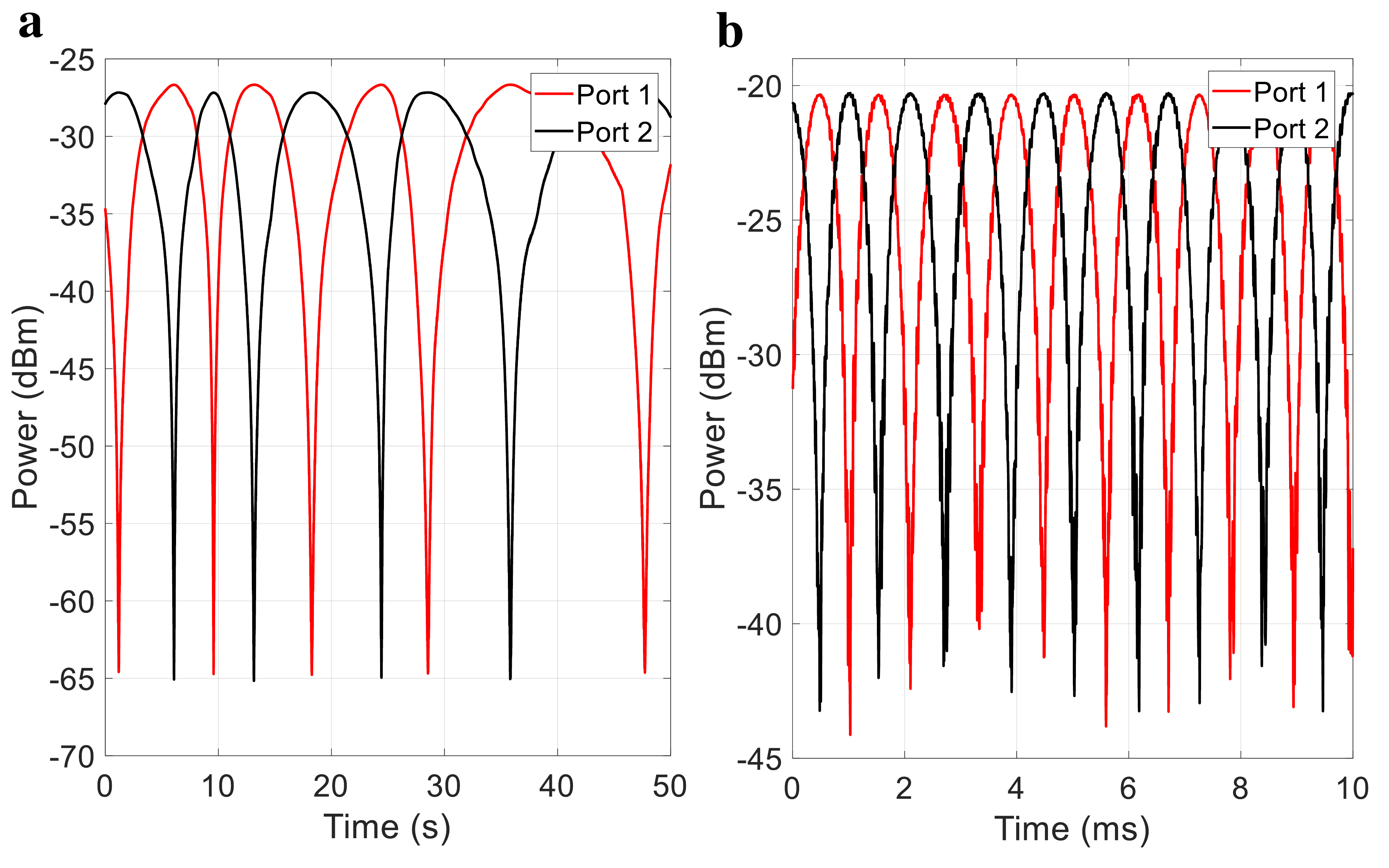}
\caption{\textbf{First order interference fringes with identical laser and two independent lasers.} }
\label{laser visibility}
\end{figure}

Here we use short fibers \zy{as the quantum channel to characterise the interference of the laser sources.}  
\zy{We first measure} the first-order interference \zy{when Alice and Bob share a common laser as their light sources and apply no encoding. Free phase drift of the short fibers allow observation of the interference fringes, as shown in Fig.~\ref{laser visibility}\textbf{a} by the optical powers measured at the interferometer output ports.
Both ports exhibit an interference fringe extinction ratio of over 35~dB, from which we obtain a near-perfect} visibility over $99.9\%$.
We then switch to pulse-carved independent lasers for Alice and Bob and use a high-speed power meter with a sampling rate of 200~kHz.  
\zy{To speed up the measurement,}  the lasers are set with a mutual frequency offset of 1~kHz. 
We record 10~s data containing 2 $\times$ $10^6$ sampling points for each port.  \zy{Figure~\ref{laser visibility}\textbf{b} shows 10~ms data segment.} 
We calculate the average visibility for each segment of 4000 samples (20~ms) from the mean values of of 5 highest and lowest data points. 
We obtain an average visibility of $98.9\%$ $\pm$ $0.1\%$. As compared to using a common laser, the visibility drops by 1~\% due to \zy{phase} noise in independent lasers and \zy{slightly temporal mismatching} between carved pulses.  

 \begin{figure}[ht]
\includegraphics[width=1\columnwidth]{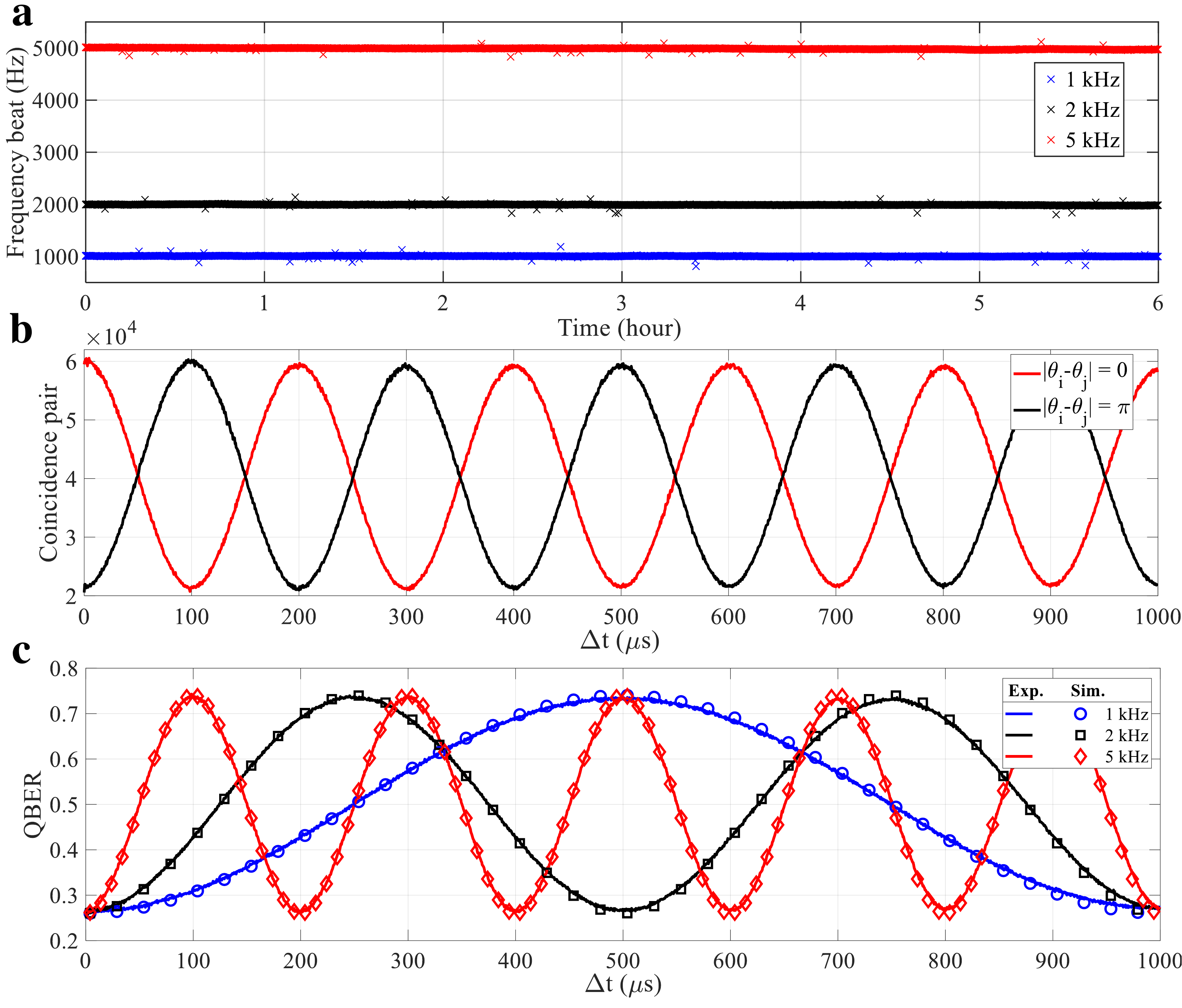}
\caption{\textbf{Characteristics of laser's stability and asynchronous two photons interference.} }
\label{freuqney difference}
\end{figure}

\zy{We then evaluate the asynchronous second-order interference
.
We prepare the lasers to have offsets of $\Delta \lzv{f}$ is set as 1, 2 and 5~kHz.
As shown in Fig.~\ref{freuqney difference}\textbf{a}, the frequency beat exhibits excellent stability over 6~h, with measured mean frequencies of 1005.2, 1991.3 and 4988.6 Hz and respective standard deviations of 5.9, 7.7 and 12.0~Hz. 
We apply alternative 0 and $\pi$ phase modulation to Alice's laser pulses while Bob's pulses are not phase modulated.  We use a superconducting nanowire single photon detector and a time-tagger to record the photon detection result. 
We then sort pair-wise coincidences according to the time separation ($\Delta t$) and the phase difference $|\theta_i - \theta_j|$ between the two time bins, and plot the count of coincidence as a function of $\Delta t$. 
Figure~\ref{laser visibility}\textbf{b} shows the interference result for $\Delta \lzv{f}$ = 5~kHz. 
For a coincidence between any two time bins, it is most (least) likely to have two photons exiting from the same output when their phase difference is 0 ($\pi$).  This phase difference includes both modulation, and evolution due to the laser frequency offset.   At $\Delta t \simeq 0$, coincidence count starts at its peak value for $|\theta_i - \theta_j| = 0$ and at its minimum for its trough for $|\theta_i - \theta_j| = \pi$, due to negligible phase evolution ($2\pi\Delta t \simeq 0$).  With further increasing time separation, we observe the coincidence count oscillates at the laser offset frequency.}
From this oscillation, we obtain an asynchronous coincidence visibility of 0.484. No deterioration of this visibility is noticed for large time-bin separation because only short fibers used in this experiment caused little decoherence. 





Based on the coincidence result, we calculate the $\boldsymbol(X)$-basis QBER $E_x$ using 
\begin{equation}
E_x = \frac{C_{\pi}}{C_0 + C_\pi},
\end{equation}
\noindent where $C_0$ ($C_\pi$) represents the coincidence rate when the same detectors clicked twice and $|\theta_i - \theta_j| = 0$ ($|\theta_i - \theta_j| = \pi$). The 5~kHz result is plotted in Fig.~\ref{laser visibility}, together with results of other frequency offsets. The experiment results \zy{are consistent} 
with our theoretical simulation. The simulation principle follows the derivation described in \zy{M}ethods of \zy{M}ain \zy{T}ext. \zy{In the absence of long fibers,} we reach a simpler equation $E_x = \frac{1}{2} - \frac{V_2}{2}{\rm cos}( 2 \pi \Delta \lzv{f} \Delta t)$ as the fiber drift is negligible compared to the frequency offset. From another perspective, the result indicates \zy{it is possible to derive $\Delta \lzv{f}$ using $[2\nu_a, 2\nu_b]$ coincidences thus to allow real-time zeroing the laser frequency offset~\cite{zhou2023quantum}}.
For $\Delta t \leq 0.5~\mu$s, we obtain $\boldsymbol{X}$-QBERs of 0.257, 0.255 and 0.255 for frequency offsets of 1, 2 and 5~kHz, respectively.  
These QBERs are close to the theoretical minimum of 0.25.


\section{Encoder}

Encoder receives continuous-wave laser input that is polarization aligned to the slow axis of its electro-optic modulators.  It is driven by an arbitrary waveform generator with a sampling rate of 25 GSa/s. Three intensity modulators are used to prepare pulses among three intensity levels, $\mu$ (signal), $\nu$ (decoy) and $o$ (vacuum). IM1 is used for pulse carving, IM2 is modulated by a waveform channel for 3-level signal intensity modulation,
and IM3 is to enhance the extinction ratio between signal and vacuum pulses. 
Two phase modulators encode the quantum signal with 16 phase slices, $\theta \in \{0, \pi/8, 2\pi/8... 15\pi/8 \}$, to meet the requirement of phase randomization in asynchronous MDI-QKD protocol. Electrically driven polarization controller is used to pre-compensate the polarization drift after transmission over the long fiber. Then VOA sets the photon flux of the quantum signal to the single photon level before launching into the quantum channel. We use a 80,000-bits pseudo-random sequence in the experiment, having a duration of 80~$\mu s$. 
Finally, our setup allows 100\% transmission duty cycle for quantum signals because strong references pulses are no longer required for our asynchronous MDI-QKD protocol. 

\section{Derivation of \texorpdfstring{$\boldsymbol{X}$}-basis QBER.}

Asynchronous two-photon interference is affected by the fiber fluctuation and the lasers’ frequency offset ($\Delta\lzv{f}$), see Fig.~3 in main text. Here we assume a stable frequency offset over the experiment time and take a Gaussian distribution for fiber drift rates~\cite{Lucamarini2018,zhou2023quantum}. 
We further assume the laser coherence time is much longer than $\Delta t$, which is the coincidence pairing interval.
$\boldsymbol{X}$-basis QBER 
has a floor error of $\frac{1-V_2}{2}$, where $V_2$ is the visibility of second order coincidence interference. The $\boldsymbol{X}$-basis QBER 
is hence derived from the phase drift characteristics
\begin{equation}
\begin{aligned}\label{eq:sim}
    E_x & = \frac{1-V_2}{2} + V_2 \int_{-\infty}^{+\infty}\frac{1-{\rm cos}(\theta)}{2}G(\omega)d \omega \\
    & = \frac{1}{2} - \frac{V_2}{2} \int_{-\infty}^{+\infty} {\rm cos}( 2 \pi \Delta \lzv{f} \Delta t + \omega \Delta t)G(\omega) d \omega\\
    & = \frac{1-V_2}{2} + \frac{V_2}{2}\left[1-e^{-\sigma^2 \Delta t^2 / 2} {\rm cos}( 2 \pi \Delta \lzv{f} \Delta t)\right],
\end{aligned}
\end{equation}
where $\omega$ is the fiber drift rate following the Gaussian distribution $G(\omega)$ with a standard deviation of $\sigma$. We then obtain a simpler formula $E_x = \frac{1}{2} - \frac{V_2}{2} e^{-\sigma^2 \Delta t^2 / 2}$ in the case for a negligible frequency offset.

\section{Drift compensation}

\begin{figure}[ht]
\includegraphics[width=1\columnwidth]{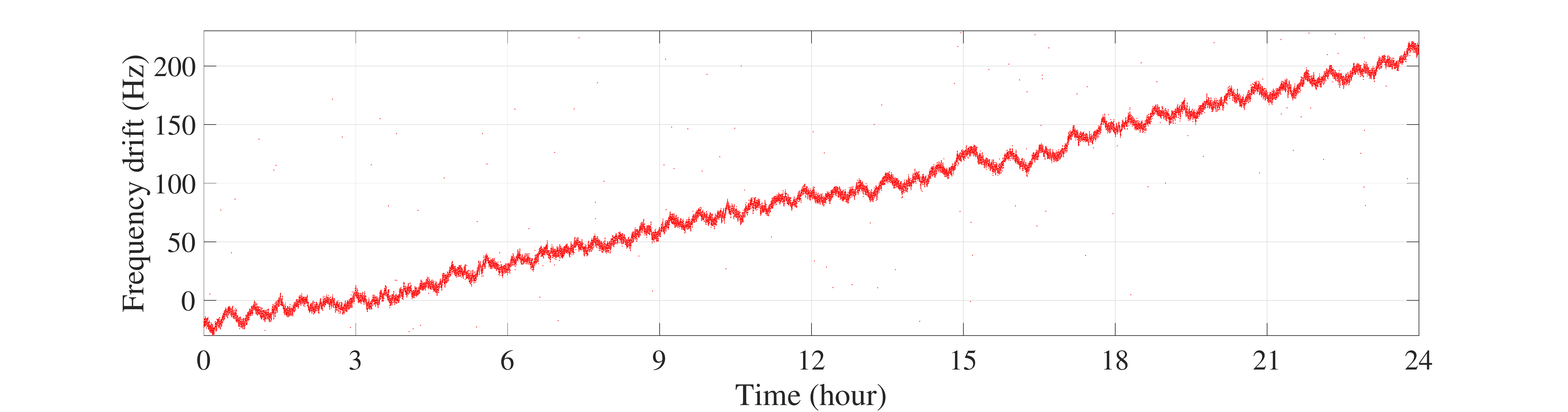}
\caption{\textbf{\lzv{Differential frequency drift between two independent lasers.}} }
\label{laser drift}
\end{figure}

\lzv{The two lasers have the same frequency drift direction, and we measure that their differential frequency drift is about 250 Hz per day, see Fig.~\ref{laser drift}. During initial setup the frequency difference ($\Delta f$) between the lasers is monitored through their beating note recorded by a fast photodiode. We can precisely set $\Delta f$ to be within 10 Hz through adjusting the modulation frequency to the electro-optical modulator (EOM) in one of the lasers. The performance of asynchronous MDI-QKD system is tolerant to phase drift from the long fiber and the small $\Delta f$. We were able to keep the two lasers free running for a full day. We recalibrate the frequency drift when $\Delta f $ exceeds a threshold value, e.g., 300 Hz.}

\lzv{Charlie uses a polarization beam splitter (PBS) at each input to ensure an identical polarization for all photons entering his 50/50 interfering beam splitter. The reflected signals from the PBS are routed to a single photon detector (monitoring polarization counts) through a polarization beam combiner. Polarization drift in the quantum link will affect the transmission through the PBS and then the detector count rate. The count rate is kept minimal via controlling an electrically driven polarization controller at the transmitter, thus maximizing the transmission to the 50/50 interfering beam splitter. This polarization compensation operates continuously at a rate of 5 - 10 Hz\cite{zhou2023quantum}. }   

\lzv{In order to obtain an optimal time alignment of two users' pulses at Charlie to ensure high visibility interference, we firstly measure the temporal misalignment between two users' pulses, and then delay the encoding pattern of one user with respect to the other. The temporal drift rate in the long fiber depends on the stability of the environment. Thanks to our air-conditioned temperature-stabilized laboratory, we make the temporal drift compensation once every 20 to 30 minutes. }

\section{System loss characterization}

\begin{table}[ht]
\caption{Lengths and corresponding losses for the fiber links in the experiments.}
\centering

\begin{tabular}{c c|c c|c c}
\hline
\hline
\multicolumn{2}{c}{Total} & \multicolumn{2}{|c}{Alice} &  \multicolumn{2}{|c}{Bob}\\
\hline
length (km) & loss (dB) & length (km) & loss (dB) & length (km) & loss (dB)\\
\hline
\hline
201.86 & 32.25 & 100.93 & 16.01 & 100.93 & 16.24\\
306.31 & 49.70 & 153.45 & 24.73 & 152.86 & 24.97\\
413.73 & 66.42 & 206.87 & 33.13 & 206.86 & 33.29\\
508.16 & 81.40 & 254.38 & 40.66 & 253.78 & 40.74\\
\hline
\hline
\end{tabular}
\label{tab:fiber_los}
\end{table}

In our setup, the quantum channel is formed by ultra-low-loss fiber spools (G654.C ULL) with a typical attenuation ranging from 0.158~dB~km$^{-1}$ to 0.162~dB~km$^{-1}$. The fiber spools are consist of 50.4 km and 25.2 km each in length, and we include 5 km conventional fiber (1 dB loss) in each side for 413 km fiber experiment. For convenience, different sets are 
combined with optical connectors, which will \zy{introduce} 
extra loss. Here we optimize fiber connections before each experiment to \zy{ensure} the quantum channel \zy{having} an average loss coefficient of 0.16~dB~km$^{-1}$. In Table~\ref{tab:fiber_los} we summarise the lengths and corresponding losses for the fiber links in the experiments. 

Charlie's module contains a series of components including polarization beam splitter, 50/50 beam splitter, manual polarization controller and the connectors, and \zy{has} a total loss of 1.50~dB before single photon detectors.

\section{Detailed experimental parameters and  results}

Table~\ref{tab:Tmean} summarises the average pairing intervals for simulation and experiment with a given $T_{c}$ in different quantum link fiber distances. 
First, the theoretical simulations reproduce almost exactly the experimental results, suggesting that the explanatory formula provided in Eq.~\eqref{T_mean} is tight. Second, \zy{there is no variation in the experimental average intervals across different coincidence sets ($\mathcal{S}[k_{a}^{\rm tot},k_{b}^{\rm tot}]$), }
which means that the pairing operation is independent of pulse intensity and the assumption of the decoy-state approach is correct in the experiment.

Table~\ref{tab:key rate} lists encoding parameters in the symmetric case. Alice and Bob share the identical set of parameters, \textit{i.e.}, $\mu_a = \mu_b =\mu$, $\nu_a = \nu_b = \nu$, $o_a = o_b = o \equiv 0$, $p_{\mu_a}=p_{\mu_b} = p_{\mu}$, $p_{\nu_a}=p_{\nu_b} = p_{\nu}$ and $p_{o_a}=p_{o_b} = p_{o}$. \zy{Weak coherent pulses} with signal ($\mu$), decoy ($\nu$) and vacuum ($o$) intensities are randomly prepared with probabilities $p_{\mu},~p_{\nu}$ and $p_o$, respectively. $n_{(k_{a}|k_{b})}$ is the number of \zy{successful clicks} announced 
by Charlie, where $k_{a}$ ($k_{b}$) is $\mu,~\nu,~o$ indicating the intensity Alice (Bob) choose. 
Note that Alice and Bob will not disclose the location of time bins with intensities $\mu$ and $o$, hence $n_{(o|\mu)}$, $n_{(\mu|o)}$, $n_{(\mu|\mu)}$ and $n_{(o|o)}$ can not be obtained in a practical implementation of asynchronous MDI-QKD. Table~\ref{tab:key rate} then summarizes the experimental results and important quantities used for key rate calculation. $n_{[k_{a}^{\rm tot},k_{b}^{\rm tot}]}$ is the successful paring number where $k_{a}^{\rm tot}$ ($k_{b}^{\rm tot}$) is the combined intensity in Alice's (Bob's) side. $m_{[\mu,\mu]}$ and $m_{[2\nu,2\nu]}$ are the total error paring numbers in the $\boldsymbol{Z}$ and $\boldsymbol{X}$ bases. $E_z$ and $E_x$ represent the error rates in the $\boldsymbol{Z}$ and $\boldsymbol{X}$ bases. The rest notations have been explained in the main text.

\begin{table}[ht]
\centering
\caption{The average paring interval at various quantum link fiber lengths.}
\setlength{\tabcolsep}{3mm}{
\begin{tabular}{c | c c c c }
\hline \hline
    Total length (km) &   201.86  & 306.31  &413.73  & 508.16  \\
 \hline
  $F $ (Hz)   & $10^{9}$ & $10^{9}$ & $10^{9}$  &  $10^{9}$  \\ 
    $T_c$ ($\mu s$) & 5 & 20 & 60 & 200\\
\hline
  Simulation  $T_{\rm{mean}}$ ($\mu s$) & 0.41 & 3.52 & 19.73 & 70.06\\
  Experiment  $T_{\rm{mean}}$ of $\mathcal{S}_{[\mu,\mu]}$ ($\mu s$) &  0.44 & 3.79 & 19.82 & 70.96
  \\
  Experiment  $T_{\rm{mean}}$ of $\mathcal{S}_{[2\mu,2\mu]}$ ($\mu s$) & 0.43 & 3.79 & 19.81 & 70.96
  \\
   Experiment  $T_{\rm{mean}}$ of $\mathcal{S}_{[2\nu,2\nu]}$ ($\mu s$) & 0.43 & 3.79 & 19.83 & 70.89 
  \\
\hline \hline
\end{tabular}
}
\label{tab:Tmean}
\end{table}

\newpage

\begin{table}[t!]
\centering
\caption{Experimental parameters and results at various quantum link fiber lengths.}
\setlength{\tabcolsep}{3mm}{
\begin{tabular}{c | c c c c }
\hline \hline
    Total length (km) &   201.86  & 306.31  &413.73  & 508.16  \\
    \hline
$\mu$ & 0.431 &0.414 &0.424 & 0.542\\
$\nu$ & 0.020& 0.024&0.030 & 0.035\\
$p_{\mu}$ & 0.252& 0.233 & 0.217 & 0.261\\
$p_{\nu}$ &  0.194 & 0.244 & 0.315 & 0.344 \\
$p_{o}$ &  0.554 & 0.523 & 0.468 &  0.395\\
 \hline
  $F $ (Hz)   & $10^{9}$ & $10^{9}$ & $10^{9}$  &  $10^{9}$  \\ 
    $N$  & $4.30\times10^{12}$ & $1.38\times10^{13}$ & $3.01\times10^{13}$ &  $7.24\times10^{13}$ \\  
    $T_c$ ($\mu s$) & 5 & 20 & 60 & 200\\
    \hline
  $n_{(\mu|\nu)}$ & 1217953802
 & 	568343320	 & 257343805	 & 173848551\\
  $n_{(\nu|\mu)}$ & 1179642539
 &  	552799554 & 	225370775 & 	172931392\\
    $n_{(\nu|\nu)}$ & 81412095
&  	65241610	 & 46086880	 & 27045205\\
  $ n_{(\nu|o)}$ & 121554019
 & 	71266320	 & 37398151	 & 15769092\\
  $ n_{(o|\nu)}$ & 117737655
	 & 69528468 & 	32182852 & 	15892169\\	 
	 \hline
    $n_{[o,o]}$ &313 & 139 & 235 &71 \\ 
    $n_{[\nu ,\nu] }$ & 1469778	& 1415687 &	1045556 & 354485\\
    $n_{[\mu ,\mu] }$ & 1092123404 &	370451795	&96538880 & 46060442\\
     $m_{[\mu ,\mu] }$ & 725019 &	223420	&107466 & 93948\\
   $n_{[\nu ,o ]}$ & 28751 &	15549	& 12990& 6269\\
   $n_{[\mu ,o] }$ & 780418 &	251029 & 125498 & 71943\\
   $n_{[o ,\nu] }$ & 27292 &	15050 &	11324 & 6269\\
   $n_{[o , \mu ]}$ & 765304 & 243558	& 108461 &71863\\ 
    $n_{[2\nu ,2\nu ]} $ &	42348 &	75628 &	113825 & 63519\\ 
   $m_{[2\nu ,2\nu] } $ &	11407  &	20680  &	 31557 & 18615 \\    $n_{[2\nu , o ]} $ & 749086 &	717129 &	600446 & 170984 \\ 
  $n_{[o , 2\nu] } $ & 702718 &	684504 &	444637 & 173734\\ 
 \hline
 $E_z$ &	0.00066 &	0.00060	 & 0.00111 & 0.00204\\
$E_x$ &	0.2694 &	0.2734	 & 0.2772 & 0.2931\\
 $\underline{s}_{11}^z$ & 460369142
& 159161908 & 39264580 & 14357572\\
 $\underline{s}_{11}^x$  &18739  &31965 &47132& 24307\\
$\overline{\phi}_z^{11}$ & 0.0916 &0.1212 & 0.1150 & 0.1960
\\
 \hline
 SKR (bit/s)	 & $ 5.7631\times 10^{4}$
 & $5.1821\times 10^{3}$
 & $5.9061\times 10^{2}$
&$42.6351$
\\
SKR (bit/clock)	 & $ 5.7631\times 10^{-5}$
 & $5.1821\times 10^{-6}$
 & $5.9061\times 10^{-7}$
&$4.2635\times 10^{-8}$
\\
$SKC_{0}$  (bit/clock) &	$8.5961\times 10^{-4}$ & $1.5459\times 10^{-5}$& $3.2898\times 10^{-7}$ & $1.0451\times 10^{-8}$\\
Ratio SKR over $SKC_{0}$&	0.0670 &	0.3352	&   1.7953
  & 4.0795\\
\hline \hline
\end{tabular}
}
\label{tab:key rate}
\end{table}

\newpage

\bibliographystyle{modified-apsrev4-2_new}
\bibliography{aa}